\documentclass[]{aa}  

\usepackage{graphicx}
\usepackage{relsize}
\usepackage{appendix}
\usepackage{txfonts}
\usepackage{gensymb}
\usepackage{siunitx}
\usepackage{microtype}
\usepackage[flushleft]{threeparttable}
\def\kms{km\,s$^{-1}$}

\begin{document}

   \title{A spectroscopic multiplicity survey of Galactic Wolf-Rayet stars \thanks{Based on observations made with the Mercator Telescope, operated on the island of La Palma by the Flemish Community, at the Spanish Observatorio del Roque de los Muchachos of the Instituto de Astrofísica de Canarias.} \thanks{Based on observations obtained with the HERMES spectrograph, which is supported by the Research Foundation - Flanders (FWO), Belgium, the Research Council of KU Leuven, Belgium, the Fonds National de la Recherche Scientifique (F.R.S.-FNRS), Belgium, the Royal Observatory of Belgium, the Observatoire de Genève, Switzerland and the Thüringer Landessternwarte Tautenburg, Germany.}}

   \subtitle{I. The northern WC sequence}

   \author{K. Dsilva
          \inst{1},
          T. Shenar\inst{1}, %\fnmsep
          H. Sana\inst{1}
          \and
          P. Marchant\inst{1}
          }

   \institute{$^1$Institute of Astronomy, KU Leuven,
              Celestijnenlaan 200D, 3001 Leuven\\
              \email{karan.dsilva@kuleuven.be}
             }

   \date{Received 25$^{\textrm{th}}$ May, 2020; accepted 13$^{\textrm{th}}$ June, 2020}

% \abstract{}{}{}{}{} 
% 5 {} token are mandatory
 
  \abstract
   {It is now well established that the majority of massive stars reside in multiple systems. However, the effect of multiplicity is not sufficiently understood, resulting in a plethora of uncertainties about the end stages of massive-star evolution. In order to investigate these uncertainties, it is useful to study massive stars just before their demise. Classical Wolf-Rayet stars represent the final end stages of stars at the upper-mass end. The multiplicity fraction of these stars was reported to be ${\sim}0.4$ in the Galaxy but no correction for observational biases has been attempted.}
   {The aim of this study is to conduct a homogeneous radial-velocity survey of a magnitude-limited ($V$ $\leq 12$) sample of Galactic Wolf-Rayet stars to derive their bias-corrected multiplicity properties. The present paper focuses on 12 northern Galactic carbon-rich (WC) Wolf-Rayet stars observable with the 1.2\,m Mercator telescope on the island of La Palma.}
   {We homogeneously measured relative radial velocities (RVs) for carbon-rich Wolf-Rayet stars using cross-correlation. Variations in the derived RVs were used to flag binary candidates. We investigated probable orbital configurations and provide a first correction of observational biases through Monte-Carlo simulations.}
   {Of the 12 northern Galactic WC stars in our sample, seven show peak-to-peak RV variations larger than 10\,\kms, which we adopt as our detection threshold. This results in an observed spectroscopic multiplicity fraction of 0.58 with a binomial error of 0.14. In our campaign, we find a clear lack of short-period (P~$<~\sim$100\,d), indicating that a large number of Galactic WC binaries likely reside in long-period systems. Finally, our simulations show that at the 10\% significance level, the intrinsic multiplicity fraction of the Galactic WC population is at least 0.72.}
   {}


   \maketitle
%
%________________________________________________________________

\section{Introduction}\label{sect:intro}

Stars with initial masses $M_i \gtrsim 8\,M_{\odot}$ are expected to end their lives as core-collapse supernovae (SNe), forming compact stellar remnants such as neutron stars and black holes (BHs) \citep{2003Heger}. Through their strong stellar winds, intense radiation and powerful explosions, they dominate the transport of energy and momentum into the interstellar medium \citep[cf.][]{2005MacLow}. A small subset of massive stars appear as Wolf-Rayet (WR) stars. They are a spectroscopic class of stars dominated by broad emission lines that are a tell-tale sign of strong, radiation driven winds \citep[see review by][]{2007Crowther}. WR stars are subdivided into three main classes: nitrogen-rich (WN), carbon-rich (WC), and the extremely rare oxygen-rich (WO) sequences. They reflect the chemical composition of their atmospheres - N-rich (products of the CNO cycle) or C/O-rich (products of He-burning). 

Due to the absence of a hydrogen envelope and their chemical composition, most WR stars are believed to be evolved, post-main sequence objects. These objects are called `classical WR stars' \citep[cf.][]{1981Massey} and they are thought to evolve spectroscopically from O-type stars to WN, WC and, finally, WO stars, possibly through an intermediate luminous blue variable or red supergiant phase \citep{1976Conti,2007Crowther}. However, very massive stars can also appear as WR stars during their main-sequence phase \citep{1997deKoter}. In contrast to the classical WRs, young WR stars contain and display hydrogen in their atmospheres. 

The role of multiplicity in the formation of WR stars is a long-standing question that is still widely debated \citep{1967Paczynski,1998Vanbeveren,2019Shenar,2020Shenar}. It is still not clear whether the dominant channel is through stripping via stellar winds or through a binary companion. This knowledge and, hence, the study of WR stars, is important to improve our understanding of observed SNe \citep{2017Zapartas,2018EldridgeCurvePop} and gravitational-wave \citep{2019Eldridge} rates, the properties of the resulting compact objects \citep[e.g.][]{2014DeMink,2016Marchant} and the evolution of massive stars \citep[e.g.][]{2003MeynetMaeder,2006Hamann,2016Shenar,2019Sander,2019Woosley}.

Large-scale multiplicity surveys in the past decade have shown that the majority of massive stars live in binaries or higher multiple systems \citep{2009Mason,2012Sana,2013Sana,2014Sana,2014Kobulnicky,2014Barba,2015Dunstall,2016MaizApellaniz,2019MaizApellaniz}. These systems are so close that the majority will interact \citep{2007Fryer,2012Sana,2013Sana,2015Dunstall}, fundamentally impacting the evolution of most of these systems \citep{1994Pols}. However, most modern radial velocity (RV) surveys of massive stars in the Galaxy have focussed on main-sequence OB-stars, with the exception of WN stars from the OWN survey \citep{2014Barba}. The state of affairs for extragalactic surveys is much better, with numerous surveys of the Magellanic Clouds \citep{1999Bartzakos,2003Foellmia,2003Foellmib,2008Schnurr} and the like for M31 and M33 \citep{2014Neugent}. 

This is not without good reason, as surveys of WR stars face numerous challenges. First and foremost, the classical WR phase is quite short-lived in the evolution of massive stars (${\sim}10\%$ of their total lifetime). Combined with the fact that most WR stars that have been discovered in the past two decades are mainly infra-red bright, the sample size for an optical RV survey is severely limited. Second, spectral lines for WR stars are formed far out in the winds. They are therefore broad and affected by variability in the winds, rendering traditional tools to measure RVs unsuitable. Additionally, the complexity of each WR system along with the diverse physics that they present (e.g. wind-variability, clumping, dust production, etc.) has prompted a system-to-system modelling, making it difficult to apply a single method in a systematic fashion to a large sample. Finally, with input on the initial conditions of the systems, recent simulations have shown that post-interaction binary systems most likely exist with periods of the order of a year \citep{2019Langer}. 

% Even in the galaxy, only a handful of hydrogen-rich WR stars are known \citep{2001vanderHucht}.
The Galactic Catalogue of WR stars\footnote{http://pacrowther.staff.shef.ac.uk/WRcat/} (henceforth GCWR) is maintained by Paul Crowther \citep[originally, Appendix 1 in][]{2015RossloweCrowther}, with 666 objects currently (v1.24). In the literature, the observed multiplicity fraction of WR stars is reported to be ${\sim}0.4$ \citep{2001vanderHucht}. This value includes both spectroscopic and visual WR binaries. However, this fraction has not been corrected for observational biases and hence the intrinsic multiplicity of WR stars is yet to be constrained. An even worse situation stands for the distributions of mass-ratios and orbital periods, which are important resources that can be used to calibrate the physics in binary stellar evolution codes (e.g. efficiency of mass-transfer and angular momentum transport) as well as what is used in stellar population synthesis \citep[e.g. BPASS,][]{2017Eldridge}. Accurate RV measurements and well-constrained uncertainties are of critical importance in order to obtain a reliable correction for observational biases. 

Binary WR candidates are usually identified through Doppler shifts, the dilution of emission lines (the companion OB-star can sometimes be more luminous in the optical), the presence of absorption lines in the spectra, photometric variability, X-ray spectroscopy for hints of RV variability and wind-wind collision, and with astrometry. Modern techniques to detect periodicities in WR stars include measuring the equivalent-width and change in the full-width half-maxima of emission lines \citep[e.g. WR1;][]{1999aMorel}. Other state-of-the-art techniques to measure RVs include measuring the centroid of the emission \citep{2012Fahed}, fitting Gaussians to line profiles \citep[e.g. WR127; ][]{2011delachevrotiere}, using (Fourier) cross-correlation \citep{2005Lefevre,2012David-Uraz,2019Shenar,2019Chene}, or even modelling excess emission \citep[e.g.][]{2000Hill}. 

The variable nature of these strong, broad emission lines of WR stars have traditionally made it challenging to detect velocity shifts of the order of a few \kms. These shifts are of the order of a fraction of a resolution element in low-resolution spectra and so they are extremely difficult to detect even with a high signal-to-noise (S/N) ratio. In terms of the parameter space for binary systems, these RV shifts correspond to systems that have long periods, and those with extreme mass-ratios. 

Clumping and time-variability studies, however, have shown that emission lines are variable on the scale of days to weeks. They have further shown that this variability is different for different ions and is mostly limited to the centers of emission lines \citep{2000LepineMoffat,2009StLouis,2011CheneStLouis}. Moreover, studies with high-resolution and high-S/N spectra have reported RV measurement accuracies of WR stars with formal uncertainties as small as ${\sim}$\SI{1}{\km\per\s} \citep[e.g. WR 113; ][]{2012David-Uraz}. This is mainly due to the fact that high-resolution spectra contain more information, and this allows for a statistically more accurate solution. Furthermore, the high-ionisation `weak' lines can also be used to derive reliable RV measurements \citep[e.g. R144, WR21a and R145; ][respectively]{2013Sana,2016Tramper,2017Shenar}. This is an advantage for systems with wind-wind collision, as the weaker lines are usually unaffected. Finally, high-resolution spectra make it possible to detect any narrow absorption lines that could be superimposed on top of the broad emission lines. 

In this context, we initiated a new monitoring program aiming at improving the observational constraints on the WR multiplicity properties. In this first paper, we present and test out RV measurement method and apply it to a magnitude-limited sample of Northern, Galactic WC stars. The sample, observations and data reduction are presented in Sect. \ref{sect:sample}. Section \ref{sect:RVdet} elaborates on the RV measurements and relevant masks used for cross-correlation. In Sect. \ref{sect:results}, the correction for observational biases is discussed, along with the spectroscopic binary fraction and intrinsic period and mass-ratio distributions. Section \ref{sect:conclusions} presents our conclusions.
%-------------------------------------------------------------------
\section{Overview of the sample and data reduction} \label{sect:sample}
%####################################################################
\subsection{The sample}
\begin{figure}
    \centering
    \includegraphics[width=9cm]{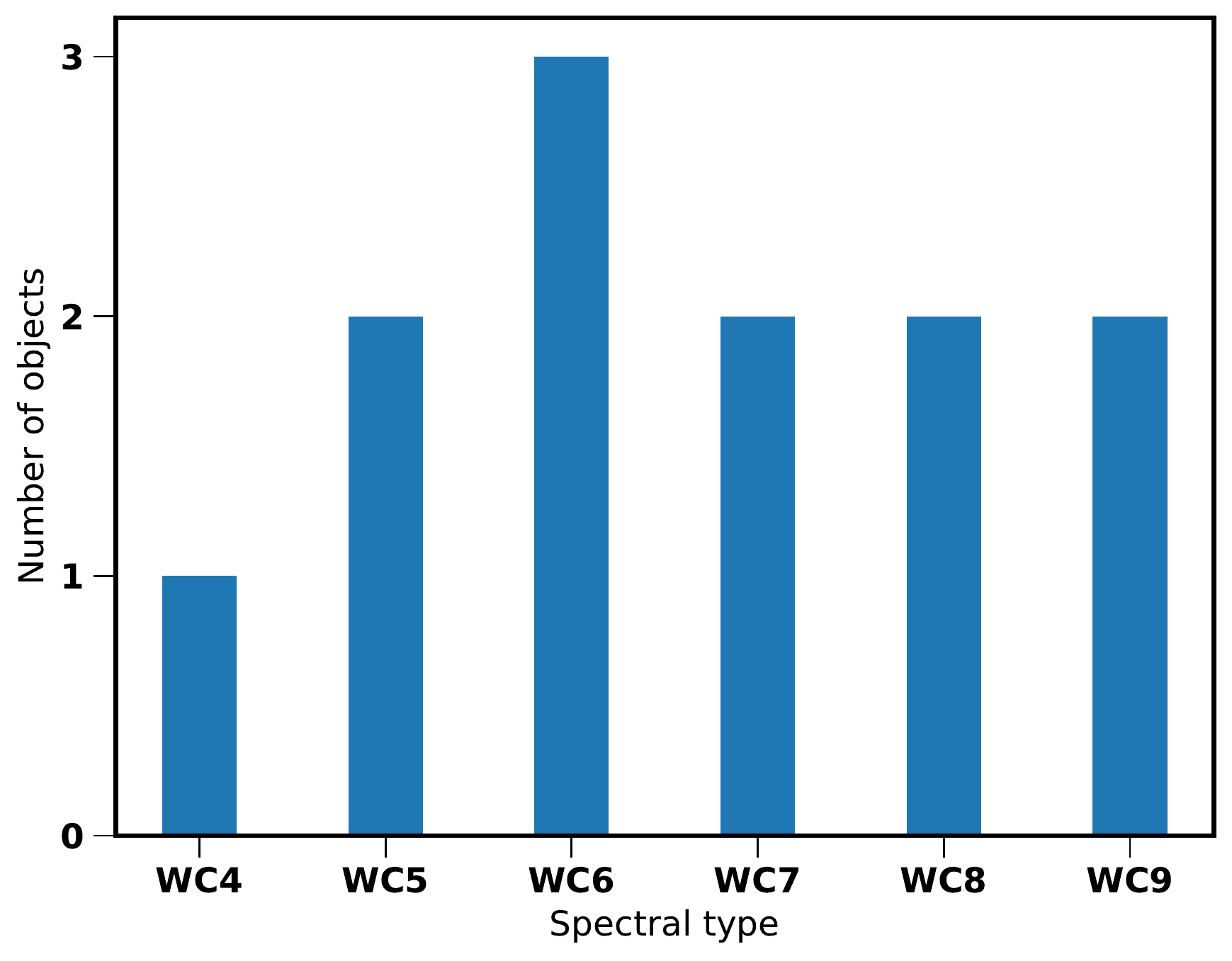}
    \caption{Spectral type distribution of the stars in our sample.}
    \label{fig:target_dist}
\end{figure}

Our sample was selected from the GCWR and contains all WC stars with $V$ $\leq$ 12 and declination $\delta \ge$ -30$\degree$ that were selected from the GCWR. For objects with missing $V$ band magnitudes, narrow-band v magnitudes \citep{1968bSmith,1984Massey} were used with the condition $v$ $\leq 13$. This resulted in a final sample of 12 WC stars. The spectral type distribution among our sample is shown in Fig. \ref{fig:target_dist}.
%####################################################################
%####################################################################
\subsection{Observations}
The WR RV monitoring campaign began in 2017 using the High-Efficiency and high-Resolution Mercator Echelle Spectrograph \citep[HERMES,][]{2011Raskin} mounted on the 1.2m Flemish Mercator Telescope on La Palma. HERMES covers the optical wavelength range from \SI{3800}{\angstrom} to \SI{9000}{\angstrom}, with a spectral resolving power of R${\sim}85000$. As we show below, the combination of high-resolution with good S/N over a long period of time results in a powerful tool to search for binarity.

For most objects, at least four epochs were obtained over the course of 15 months. Table \ref{tab:epochs} gives the an oerview of the campaign (including archival HERMES data). The number of epochs for WR 137 is larger because we obtained an additional 16 epochs with a high cadence over 15 days, with three observations over the same night to investigate the effect of short-term atmospheric/wind variability on the measured RVs (Sect. \ref{sect:wind-var}). 
%####################################################################
\begin{table}
\centering
\caption{Sample of WC stars in the RV monitoring campaign with the number of spectra, time coverage, and average S/N per resolution element at \SI{5200}{\angstrom}}
\begin{tabular}{cccc}
\hline \hline
WR\# & Spectra & Time coverage (d) & S/N @ \SI{5200}{\angstrom} \\ \hline 
4 & 22 & 890.65 & 62 \\ 
5 & 7 & 718.20 & 59 \\ 
111 & 8 & 741.94 & 157 \\ 
113 & 9 & 727.85 & 116 \\ 
117 & 4 & 487.73 & 7 \\ 
119 & 7 & 820.70 & 32 \\ 
135 & 16 & 2659.71 & 166 \\
137 & 28 & 2653.77 & 175 \\
140 & 13 & 2680.68 & 203 \\
143 & 6 & 75.96 & 33 \\ 
146 & 8 & 846.72 & 8 \\
154 & 4 & 23.94 & 48 \\ \hline
\end{tabular}
\label{tab:epochs}
\end{table}
%####################################################################

\subsection{Data reduction and normalisation}
The HERMES spectra were first reduced by the HERMES pipeline which includes the bias, flat-field corrections, cosmic removal and order merging of the spectra \citep{2011Raskin}. However, the resulting spectra are still affected by atmospheric extinction, telluric lines, interstellar reddening and the instrumental response. In order to achieve a normalisation procedure that is homogeneous over multiple epochs, we developed an algorithm that corrects for these effects. This minimises the effect of variable normalisation on the RV measurements.

We first correct the spectra for atmospheric extinction, applying a wavelength and airmass-dependent correction. Molecfit \citep{2015SmetteMolecfit,2015KauschMolecfit} is then used to remove telluric lines with meteorological conditions as input from the Mercator meteo-station. Standard calibration stars that are used for the order merging are then used to derive the instrumental response function for the night. Once corrected for these three effects, the only remaining effect is due to that of interstellar (or even circumstellar) reddening.

\begin{figure*}
    \centering
    \includegraphics[scale=0.9]{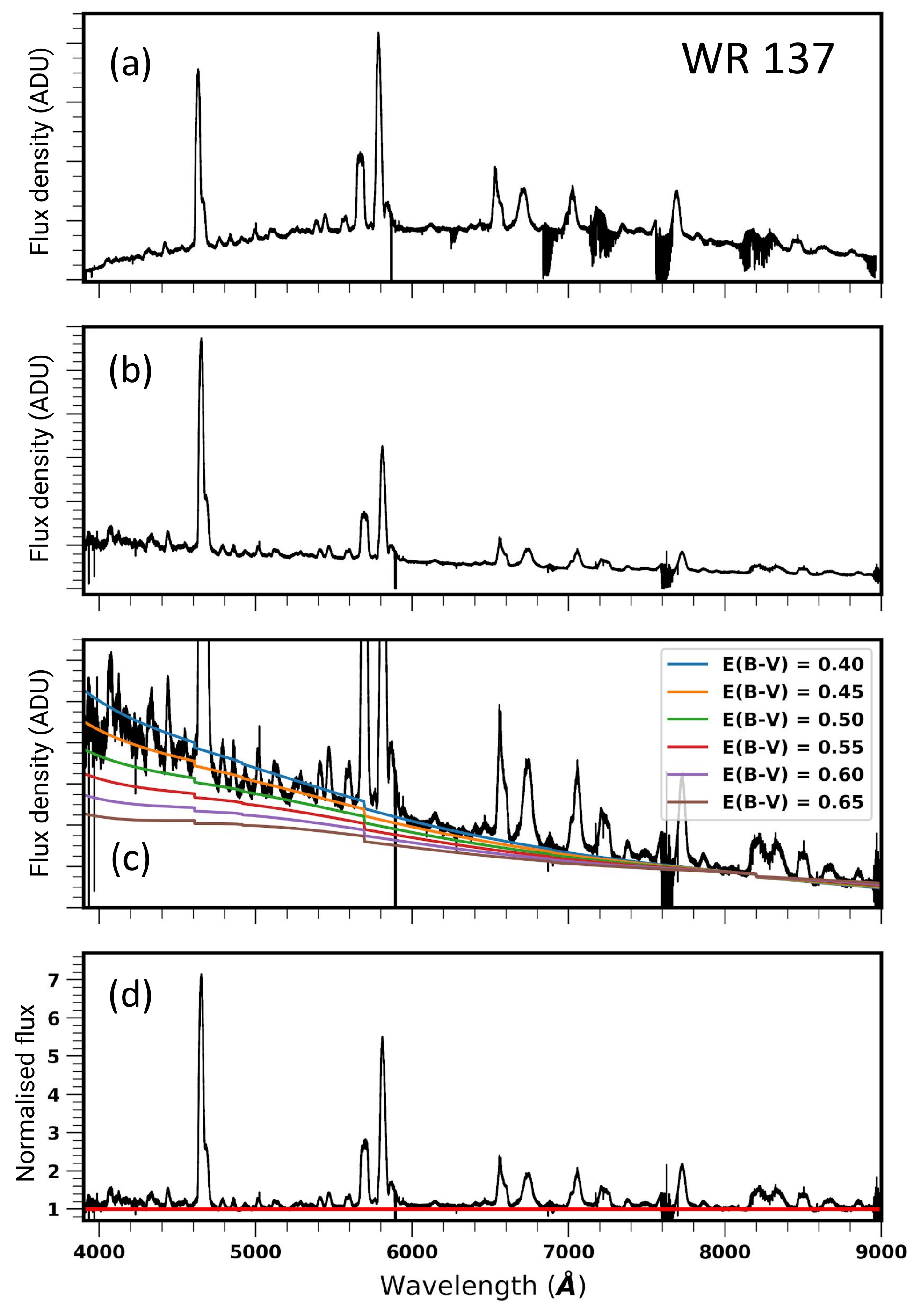}
    \caption{Main steps in our data reduction process. From top to bottom: (a) Flatfield corrected spectrum produced by the HERMES pipeline. (b) Spectrum corrected for the instrumental response function and telluric contamination. (c) Continuum flux models from PoWR to the pseudo-continuum region around \SI{5200}{\angstrom} for different values of E($B-V$). (d) Resulting normalised spectrum.}
    \label{fig:custom_pipeline}
\end{figure*}

To normalise these spectra, pseudo-continuum regions were chosen based on normalised model spectra for WC stars from the Potsdam Wolf-Rayet code \citep[PoWR:][]{2002Grafener,2003HamannGrafenerPoWR,2012Sander}. Two continuum regions were adopted: in the red, around \SI{8100}{\angstrom}, and in the blue, around \SI{5200}{\angstrom}. By anchoring the respective continuum models to the red point and then reddening them based on \citet{2004Fitzpatrick} we fit the slope and normalise the spectra. For this purpose, we used an R$_V$ value of 3.1, the average value for the Galaxy. The various steps in the data handling are illustrated in Fig. \ref{fig:custom_pipeline}. We note that we only rely on the above-mentioned reddening law to obtain a homogeneous normalisation process and therefore refrain from providing the reddening values obtained though the continuum-fitting procedure.

%-------------------------------------------------------------------
\section{Radial velocity determination}
\label{sect:RVdet}

In order to derive RVs for broad WR emission lines in a systematic manner, one needs a method that can be applied to most, if not all, of the objects in our sample and that yields meaningful errors. The uncertainty on each measurement will have contributions from different sources: statistical (based on the method and S/N), systematic (instrumental and normalisation), and intrinsic (variability in spectral lines). It is important to have reliable estimates of the uncertainties affecting the RV measurements as these will directly propagate into the observational bias correction. For binary systems, an upper limit on the RV measurement uncertainties can be estimated by measuring the dispersion in the residuals after fitting an orbit. However, the very long periods involved do not allow us to use this. 

In this study, we focus solely on RV measurements of WR stars. In the case of SB2 binaries (i.e. double-lined spectroscopic binaries), it is, in principle, also possible to measure the RVs of the secondary star (typically an OB-type star). However, as the goal of our current study is binary detection rather than orbital analysis, there is little advantage in doing so. The WR component, being typically the lighter component, would exhibit larger RV amplitudes than its OB-type companion. Moreover, applying additional binary-criteria to SB2 binaries would require us to introduce non-trivial adjustments to our bias correction (see Sect. \ref{subs:intMulProp}). 

In all cases, the estimation of uncertainties depends on the method used to measure the RVs. In the case of Gaussian fitting, the errors on the RV measurements are usually derived using the co-variance matrix of the least-squares fit. Cross-correlation relies on applying a peak-finding algorithm to the cross-correlation function (CCF), which can be a Gaussian, Lorentzian, parabola or, a sinc function \citep[e.g.][]{1993TodyIRAF}. Another approach is to find the bisector of the CCF. The errors are derived based on the width, number of pixels, height of the CCF peak, and the antisymmetric noise \citep{1979TonryErrors}. When measured for WR stars with low-resolution spectra, the errors are magnified by the low number of pixels around a broad CCF peak. In this work, we use a different statistical formulation for the CCF that represents it as a log-likelihood function. The latter is then maximised and fit with a parabola. This is explained in detail below.
%#################################################################### 
\subsection{Cross-correlation}
The method that we used for the cross-correlation is based on \citet{2003Zucker}. The method, assumptions and limitations are briefly discussed here. For a more thorough derivation we refer to \citet{2003Zucker}. The observed spectrum is denoted by $f(n)$ and is assumed to be Doppler shifted with respect to the `template' of zero-shift, denoted by $g(n)$. Both the spectrum and the template are described as a function of pixel, $n$, such that a Doppler shift results in a uniform, linear shift in the spectrum \citep{1979TonryErrors}. 

In this work, the observed spectrum is assumed to be reproducible by scaling the template with a constant ($a_0$) and shifting it by a certain number of pixels ($p_0$). The noise in the observed spectrum is assumed to be Gaussian with a fixed standard deviation ($\sigma_0$), so that:
\begin{equation}
f(n) = a_0 g(n-p_0) + d_n \textrm{, with}
\end{equation}
\begin{equation}
d_n \sim N(0,\sigma^2) \textrm{.}
\end{equation}
The spectra (or specific masks) are assumed to be `continuum-subtracted', such that they have a zero mean:
\begin{equation}
\sum_n f(n) = 0 \textrm{, and}
\end{equation}
\begin{equation}
\sum_n g(n) = 0 \textrm{.}
\end{equation}

The last assumption that the number of pixels, $N$, is constant breaks down. For very small velocity shifts, the edge effects are negligible. However, for velocities of the order of 1000 km\,s$^{-1}$, the assumption is not applicable and hence, we use a rolling function to carry out the shift in pixels. This does not have any impact on the computation of the CCF or the determination of the peak. 

In order to express the CCF in terms of a log-likelihood function, for a given shift of $p$ pixels we first define the cross-covariance function,
\begin{equation}
R(p) =\frac{1}{N}\sum_n f(n) g(n-p)\textrm{.}
\end{equation}
The standard deviation of the spectrum ($s_f$) and template ($s_g$) can be expressed as 
\begin{equation}
s_f^2 = \frac{1}{N} \sum_n f^2 (n) \textrm{, and}
\end{equation}
\begin{equation}
s_g^2 = \frac{1}{N} \sum_n g^2 (n) \textrm{.}
\end{equation}
Following this, we define the CCF of $f$ and $g$ as
\begin{equation}
C(p) = \frac{R(p)}{s_f s_g} \textrm{.}
\end{equation}
This CCF can be written in the form of a log-likelihood function. We denote a shift $\hat{p}$ that maximises the CCF. The errors can then be derived from the covariance matrix following maximum log-likelihood theory and can be expressed as
\begin{equation}
\sigma_p^2 = - \left[N \frac{C''(\hat{p})}{C(\hat{p})} \frac{C^2(\hat{p})}{1 - C^2(\hat{p})} \right]^{-1} \textrm{,}
\label{eq:errorCCF}
\end{equation}
where $C''(p)$ is the second derivative of the CCF. Equation \ref{eq:errorCCF} shows that the errors depend on the number of pixels $N$, the sharpness of the peak $C''(\hat{p}) / C(\hat{p})$ and the line S/N ratio $C^2(\hat{p}) / (1 - C^2(\hat{p}))$. 

The success of the cross-correlation depends on the accuracy and stability of the template. This is due to the fact that the template is implicitly assumed to represent the data accurately. In practice, the template is chosen to be the observations with the highest S/N. A parabola is fit to the top of the CCF in order to determine the peak and the related derivatives. The errors from the least-squares fitting of the parabola agree with those derived from Eq. \ref{eq:errorCCF}, giving confidence in the validity of the method. Once the RVs and measurement uncertainties are derived, a weighted average of the shifted spectra is used to derive a co-added spectrum that has a higher S/N. This is then used as a template to re-derive the relative RVs and, thus, in an iterative fashion, an ideal mask and a final measurement are converged upon. We note that the RVs derived are relative to a particular observation in time, and that the velocities measured are, by no means, absolute values. In order to compare with those in the literature or to the rest frame, a uniform shift must be imposed. 

%####################################################################
\begin{figure}
    \centering
    \includegraphics[width=9cm]{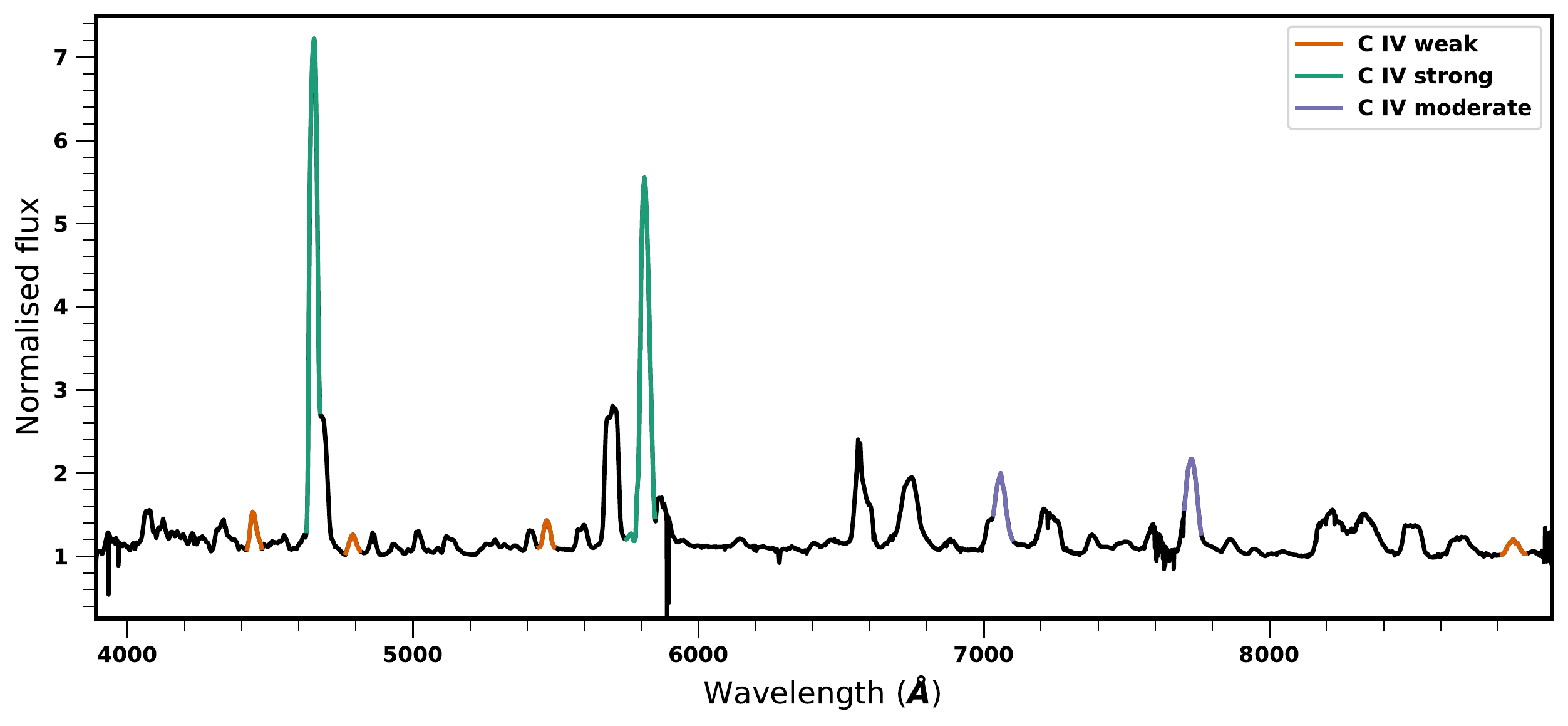}
    \caption{Spectrum of WR 137, rebinned for clarity, with different masks of C IV indicated (see legend). This process is repeated for all other ions, producing a plethora of masks.}
    \label{fig:line_strengths}
\end{figure}
%####################################################################
\subsection{Mask selection}
After addressing the choice of the template, the question of which spectral lines or wavelength ranges (henceforth `masks') to be used for the RV measurements arises. Most traditional methods of measuring RVs are applied to the strongest lines, as they contain the most signal (see Sect. \ref{sect:intro}). 
On the other hand, different lines originate in different ions and form in different regions in the wind. Therefore, by comparing the RVs measured using different sets of lines, we can both probe different parts of the wind as well as identify the most reliable mask to represent the motion of the star.

However, with high-resolution and good S/N, weak lines also contain significant amounts of information that can be used. As the WR spectra contains much broader (and stronger) spectral features, using the entire spectrum in the cross-correlation process can mitigate the problem of contamination from the secondary \citep{2012David-Uraz}. To investigate which masks are the most appropriate and if individual lines are better than the full spectrum, RVs are first measured using a variety of masks. These masks are created by selecting lines of different strengths and different ions, as well as with the full spectrum.

For WC stars, most lines originate from carbon (C II, C III and C IV), oxygen (O II, O III, O V, O VI) and helium (He I, He II). Naturally, the strength of the lines of various ions depends on the temperature (among other things) of the WR star, and are identified star-by-star using PoWR models with parameters from \citealt{2012Sander}. For each star in the sample, this results in `strong', `moderate' and `weak' masks for different ions. An example of C IV masks for WR 137 is shown in Fig. \ref{fig:line_strengths}.

The motivation for using lines of the same ion in the same mask is straightforward, while that for using only strong or weak lines is a bit more complex. Lines of the same ion are thought to have formed in the same region of the wind and should thus provide the same RV measurement. Moreover, it is useful to use only strong lines to measure RVs in order to compare the results of this work to what is available in the literature. The errors from the CCF (Eq. \ref{eq:errorCCF}), additionally, depend inversely on the S/N in the line, which is higher for strong lines. However, weak lines of highly ionised ions are formed closer to the surface and thus are thought to be less susceptible to wind variability, making them a logical choice for a mask. It is further essential to isolate and quantify the effects of wind variability on WR spectra. Once this is accounted for, we can decide which masks are the most appropriate to detect genuine RV variability. This is addressed in the following section.
%####################################################################

\subsection{Uncertainties from systematic sources and wind-variability: A proto-study of the binary WR 137} \label{sect:wind-var}
In this section, we investigate the effect of systematic sources and wind variability on RV measurements. To understand the extent to which wind variability can affect our measured RVs, WR 137 (WC7pd + O9 V) is an ideal candidate to examine. It is a well-studied long-period binary system with an orbital period of 13.05 years \citep{2005Lefevre,2016Richardson}. We obtained 16 spectra over the course of 15 days during August 2019, significantly away from its periastron passage (which will happen in 2022). Therefore, we can neglect the effects of orbital motion and wind-wind collision \citep[e.g.][]{2002Luehrs} due to binary interaction on the RV measurements for the purpose of this short-cadence study.

\begin{figure}
    \centering
    \includegraphics[width=9cm]{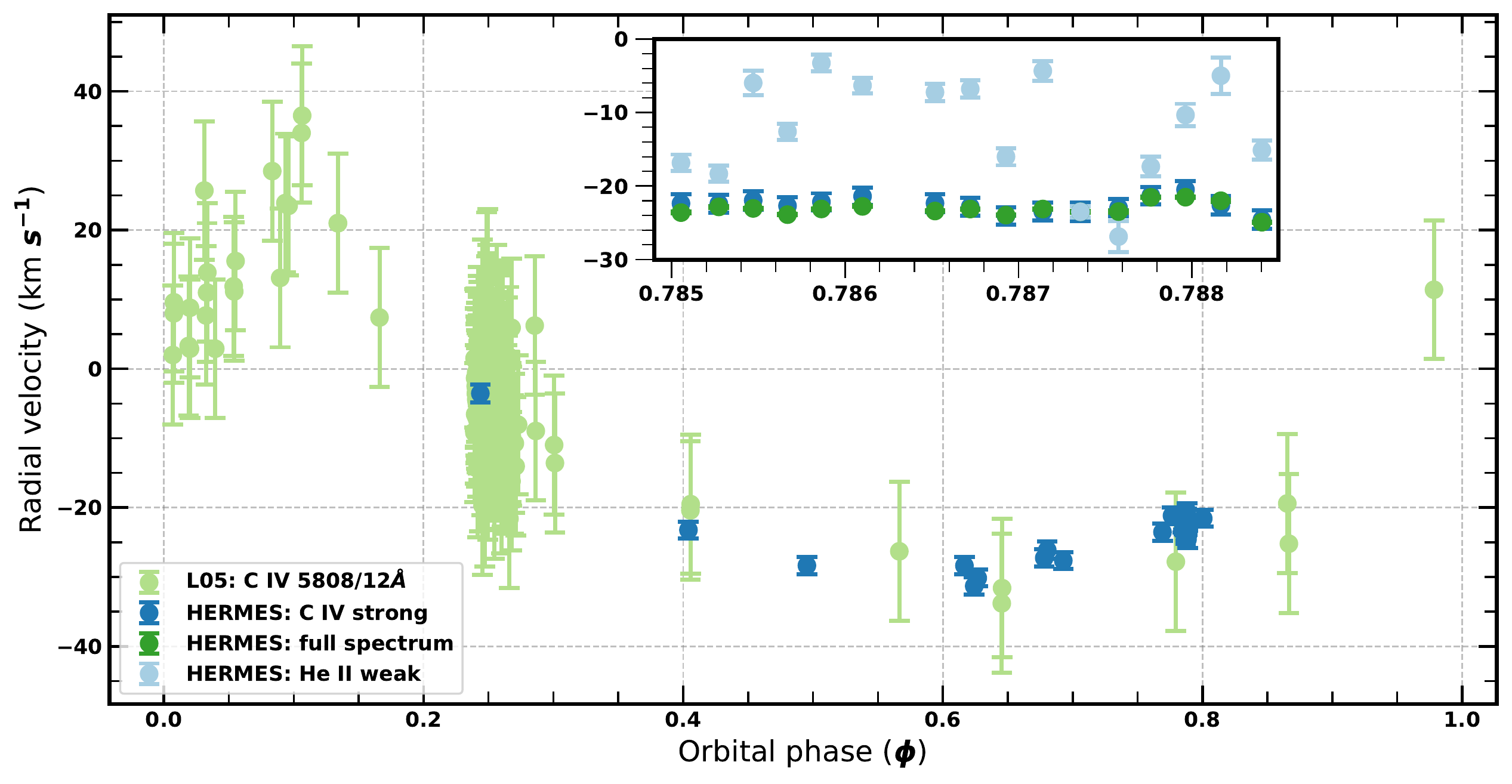}
    \caption{Phased RV plot for WR 137. Archival RV measurements are shown in blue and are taken from \citet{2005Lefevre}. The RV measurements from HERMES using the strong C IV lines are shown in red. Along with these measurements, the inset shows the short cadence measurements with the full spectrum (green) and weak He II lines (purple). All HERMES RV measurements have been shifted by $-$23.5~\kms\,in order to convert relative RV measurements to absolute values.} 
    \label{fig:WR137_shortcadence}
\end{figure}

\begin{figure*}
    \centering
    \includegraphics[width=\textwidth]{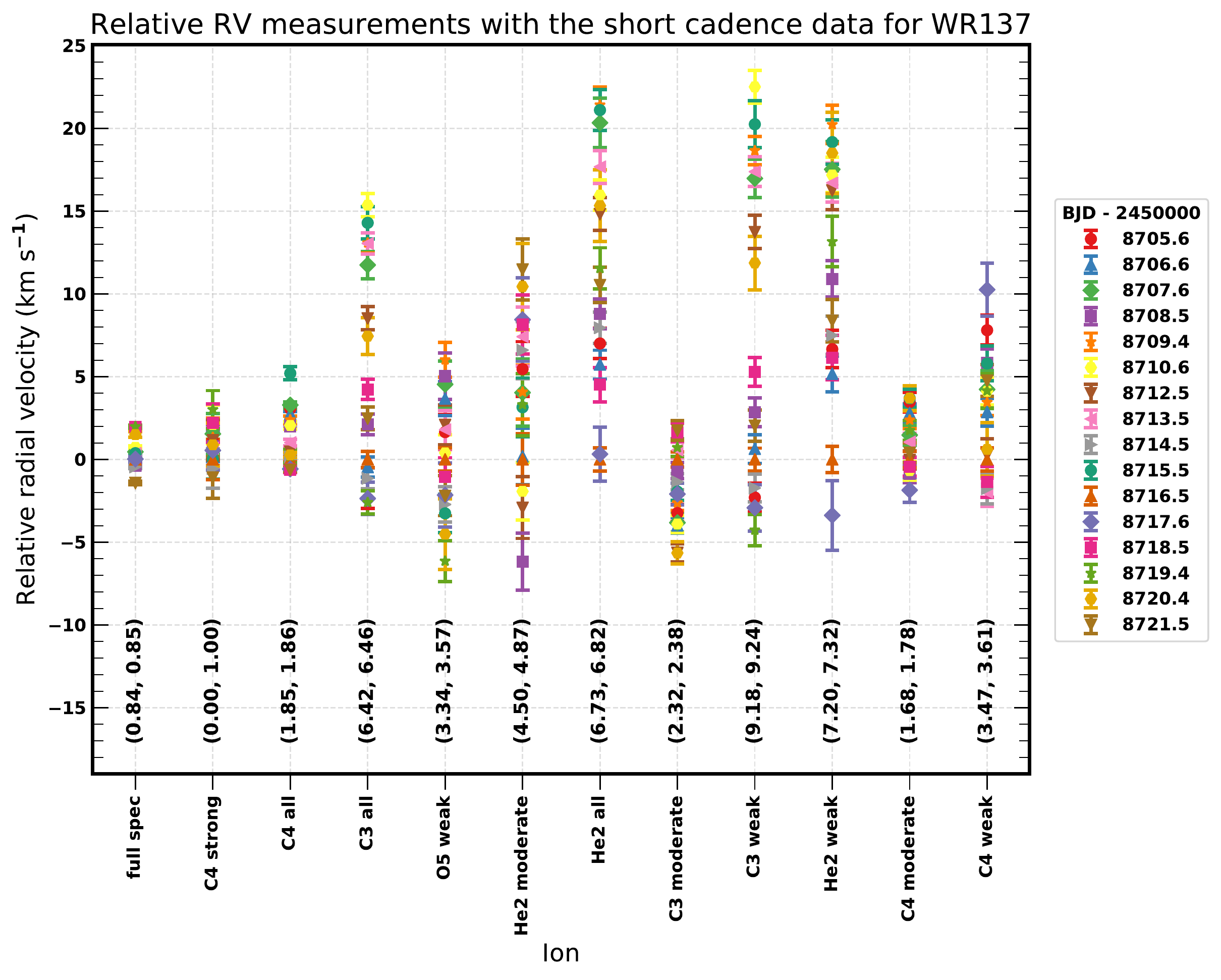}
    \caption{RV measurements for WR 137 during the short cadence study. At the bottom of the plot are the values of ($\sigma_w$, $\sigma_{RV}$) (\kms) for each mask. It is observed that the helium lines (and weak lines in general) have large dispersions. The masks least affected by wind variability are `full spec', `C4 strong', `C4 moderate', and `C4 all' (determined by their values of $\sigma_w$).} 
    \label{fig:ions}
\end{figure*}

The RVs were measured as described in the previous subsection with different masks, with the additional condition that lines contaminated by absorption from the companion were excluded. The RV measurements using different masks with strong C IV lines, weak He II lines and the full spectrum can be seen in Fig. \ref{fig:WR137_shortcadence}. Given RV measurements ($\varv_i \pm \sigma_i$), we calculated the weighted mean ($\mu_{RV}$) and standard deviation ($\sigma_{RV}$) as:
\begin{equation}
w_i = \frac{1}{\sigma_i^2}  \textrm{,}
\end{equation}
\begin{equation}
\mu_{RV} =  \frac{\sum_i w_i \varv_i}{\sum_i w_i} \textrm{,}
\end{equation}
\begin{equation}
\sigma_{RV} = \left( \frac{\sum_i w_i (\varv_i - \mu_{RV})^2}{\sum_i w_i} \right )^{\frac{1}{2}} \textrm{,}
\end{equation}

where $\sigma_{RV}$ is a measure of the real dispersion in the data. Since we have already ruled out orbital and wind-wind collision effects, this dispersion can be assumed to represent the statistical error ($\sigma_p$) along with some contribution from systematic sources and wind variability ($\sigma_w$). These errors are not expected to be correlated and are therefore added in quadrature, such that they can be expressed as follows:

\begin{equation}\label{eq:sig_RV}
\sigma_{RV} = \sqrt{\sigma_p^2 + \sigma_w^2} \textrm{.}
\end{equation}
%####################################################################
%####################################################################
Accounting for the wind variability provides us with an estimate on the effective uncertainty on the RV measurements ($\sigma_{RV}$). As $\sigma_p$ is already derived, $\sigma_w$ can be calculated for different ions and different line strengths, giving us a direct probe of how wind variability affects these lines. For now, however, it is sufficient to inspect $\sigma_{RV}$ for different ions. Figure \ref{fig:ions} depicts this comprehensively, plotting the RVs measured by each mask over the short cadence period, with $\sigma_{RV}$ indicated at the bottom of the plot.

Upon examination, two observations can be made based on Fig. \ref{fig:ions}. The first is that for masks with He II lines, a large scatter is seen independent of the strength of the lines. The second is that across other ions, masks with strong lines perform better than those with weak lines. It is not trivial to explain these observations, as there are multiple factors that are at play for each line (e.g. oscillator strength, number populations of the particular state, line-formation region, etc.).

% This is not surprising, as ions with a lower ionisation state are formed in cooler regions of the wind, which are farther out and are hence more affected by wind variability. However, it may seem counter-intuitive that weak lines are also susceptible to wind variability, but it is easy to explain. The formation region for weak lines is quite a thin shell as compared to the more extended one for stronger lines. As a result, a small perturbation of the wind is more significant in a thin shell than a thick one, resulting in a larger effect on the measured RVs. This reasoning is strengthened by the fact that weak lines of highly ionised ions (e.g. C IV weak) perform better than weak He II lines. 

In order to ascertain that the statistical errors do not bias our mask selection, we calculated values for $\sigma_w$ using Eq.~\ref{eq:sig_RV}. The results are twofold - masks with ions of a higher ionisation state are the least affected by wind variability and strong lines are affected less than weak lines. Finally, by comparing the absolute values of $\sigma_w$ in Fig. \ref{fig:ions}, the full spectrum mask is least affected by wind variability. For strong C IV lines, the average $\sigma_p$ is larger than the corresponding value of $\sigma_{RV}$, implying that the wind variability is within the statistical error (and hence $\sigma_w=0$). Given that the winds of all WC stars do not necessarily exhibit the same characteristics, this exercise has to be conducted for each object individually in order to select a specific ion to use. Upon inspection, diagnostic plots such as Fig. \ref{fig:ions} for each object indicate that the full spectrum shows the least scatter. Considering the fact that the WR component dominates the spectrum for all of our objects, we decided to use the full spectrum to measure RVs.

%####################################################################
%####################################################################
%-------------------------------------------------------------------
\section{Results and discussion}\label{sect:results}
%####################################################################
\subsection{Observed spectroscopic binary fraction}
For all the objects in our sample, RVs were measured using multiple masks. The tests described in the previous section reveal that the measurements using the full spectrum are the most reliable and we adopt those in the following section. These are given in App. \ref{s:tables_RV}. To identify objects with significant RV variation, we look at the maximum RV variation for each star, given by 
\begin{equation}\label{eq:del_RV}
\Delta \textrm{RV} = \textrm{max}(|\varv_i - \varv_j|),
\end{equation}
where $\varv_i$ and $\varv_j$ are the RV measurements for two epochs $i-j$ for a given star. One then imposes a minimum amplitude threshold $C$ for RV variations computed from any individual pair of epochs to be `significant'. 
\begin{equation}\label{eq:signif_crit}
\Delta \textrm{RV} \ge C.
\end{equation}

Unlike in \citet{2013Sana}, we do not enforce a significance criteria for each pair of RV measurements (Eq. 4 in \citealt{2013Sana} compared to Eq. \ref{eq:signif_crit} above). This is due to the fact that the values for $\sigma_p$ are typically very small (sub-\kms) and underestimate the true error (see Sect.\,\ref{sect:wind-var}). 

\begin{figure}
\centering
\includegraphics[width=9cm]{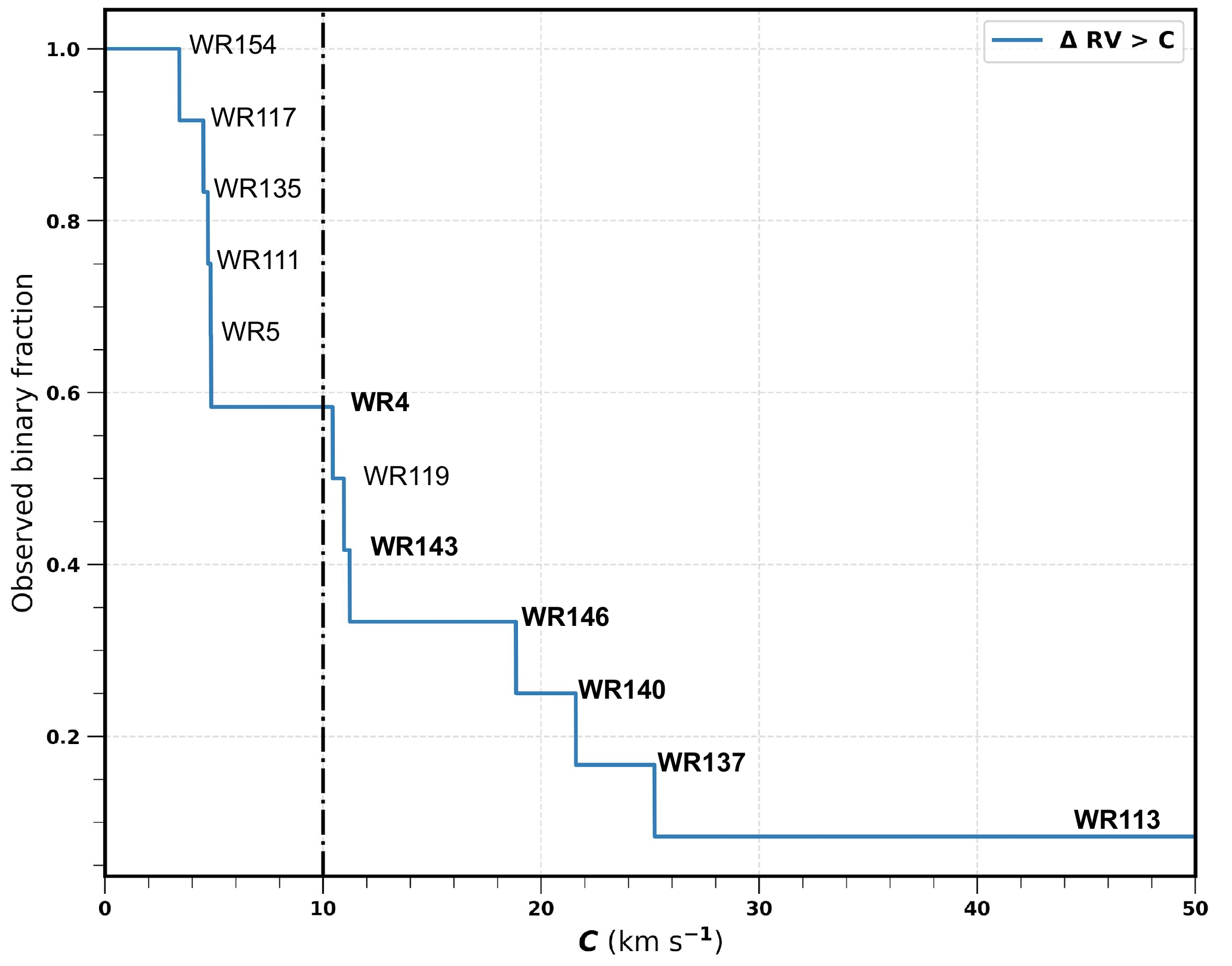}
\caption{Observed binary fraction ($f_{obs}$) as a function of the threshold $C$. The adopted threshold is the vertical dotted-dashed line at \SI{10}{\km\per\s}, resulting in $f_{obs}$=0.58. The objects are labelled as and when they are classified as binaries based on $\Delta$ RV. Names in bold are known binaries.}
\label{fig:bin_frac}
\end{figure}

Given this criterion, the observed fraction of RV variable objects ($f_\mathrm{obs}$) can be calculated as a function of threshold $C$ (Fig. \ref{fig:bin_frac}). It is encouraging to see all the known binary systems group together on the right side of the plot. We can now take advantage of the fact that a subset of our sample are confirmed binaries to constrain a suitable value for $C$.

Two kinks can be seen, in the ranges of 5 to 10\,\kms~ and of 12 to 19\,\kms~. A threshold value of 10\,\kms~ leads to an observed binary fraction of $f_\mathrm{obs}$ = 0.58, and that of 15\,\kms~ results in \mbox{$f_\mathrm{obs}$ = 0.33}. To include the known and candidate binaries in our binary fraction, we choose the former threshold of \SI{10}{\km\per\s}. 

%####################################################################
\begin{table*}
\centering
\caption{Results of the RV measurements for the sample of WC stars. $\Delta$ RV (Eq. \ref{eq:del_RV}) and $\sigma_{RV}$ (Eq. \ref{eq:sig_RV}) are calculated in this work with the full spectrum as a mask and are used to classify the objects as binaries or single stars. We do not include the archival period of 2.4096~d for WR 4 due to the fact that we rule out short-scale variations (Appendix \ref{apdx:comments}). RV plots for the known binaries can be found in Figs. \ref{fig:WR137_shortcadence} and \ref{fig:known_orbits}.}

\begin{threeparttable}
\begin{tabular}{ccccccccc}
\hline \hline

WR\# & Spectral Type & \multicolumn{2}{c}{Binary Status}  & Period & e & $\Delta$ RV  & $\sigma_{RV}$  \\ \cline{3-4}
 & (GCWR) & GCWR & This work & (d) & &(\SI{}{\km\per\s}) &(\SI{}{\km\per\s})\\\hline 
4 & WC5+? & SB1, no d.e.l. & binary & - & - & 10.4 & 2.5\\
5 & WC6 & - & single & - & - & 4.9 & 1.7\\
111 & WC5 & - & single & - & - & 4.9 & 1.6\\
113 & WC8d+O8-9IV & SB2 & binary & 29.700 $\pm$ 0.001\tnote{(a)} & 0 (fixed)\tnote{(a)} & 282.6 & 93.9\\
117 & WC9d & - & single & - & - & 4.5 & 1.5\\
119 & WC9d & - & binary & - & - & 11.0 & 3.2\\
135 & WC8 & - & single & - & - & 4.7 & 1.6\\
137 & WC7pd+O9 & SB2 & binary & 4766 $\pm$ 66\tnote{(b)} & 0.18 $\pm$ 0.04\tnote{(b)} & 25.2 & 4.6\\
140 & WC7pd+O4-5 & SB2, VB & binary & 2895.68 $\pm$ 0.17\tnote{(c)} & 0.8996$^{+0.0008}_{-0.0008}$\tnote{(c)} & 21.6 & 6.5\\
143 & WC4+Be & d.e.l. & binary & - & - & 11.2 & 4.0\\
146 & WC6+O8 & SB2, VB & binary & - & - & 18.9 & 5.9\\
154 & WC6 & - & single & - & - & 3.4 & 1.3\\ \hline 
\end{tabular}
\begin{tablenotes}
    \item[(a)] \citet{2012David-Uraz}
    \item[(b)] \citet{2005Lefevre}
    \item[(c)] Thomas et al. submitted
\end{tablenotes}
\end{threeparttable}
\label{tab:WC_data}
\end{table*}
\footnotetext[2]{David-Uraz}
%####################################################################
%####################################################################

%-------------------------------------------------------------------
\subsection{Comparison with the literature}\label{subsec:GWC_bin}
Although the magnitude-limited sample of WC stars in this study is small, the even distribution across spectral types (Fig. \ref{fig:target_dist}) omits any selection bias. With this in mind, two important conclusions can be drawn from Fig. \ref{fig:bin_frac}. The first is that our value of $f_{\textrm{obs}}$ is comparable to the value of 0.56 observed for main-sequence O-type stars by \citet{2012Sana}.

The second is that there seems to be a lack of short-period (P $<$ 100~d) binaries compared to the overabundance of short-period systems observed for main-sequence O-type stars \citep{2012Sana}. This is especially surprising given that the observational bias favours the detection of short-period WR binaries, which are expected to exhibit high RV amplitudes. Of the known binaries in our sample, orbits are derived only for WR 137, WR 140 and WR 113, with only WR 113 being part of a short-period binary system (P $\sim$ 29.7~d). This low fraction of detected short-period binary systems (0.08) is in stark contrast to what has been found for main-sequence O-type stars.

In order to investigate the observed lack of short-period WC binary systems, we scoured the literature in order to expand our small sample. The GCWR reports 237 WC stars in the Galaxy, of which 87 were included in \citet{2001vanderHucht}. For the sake of considering WC stars that were thoroughly studied for binarity, we focus on this subset of 87 stars. 

Of this subset, 37 WCs are classified as detected or probable binaries, resulting in an observed Galactic WC binary fraction of 0.41. However, this fraction is severely biased towards short-period systems. This is because historical studies were mainly sensitive to short-period systems, given typical RV measurement uncertainties of the order of ${\sim}$20 km\,s$^{-1}$. Despite the large bias towards detecting short-period systems, 15 report periods shorter than 100~d, seven of which have derived orbits. Therefore, the fraction of Galactic WC stars that are conclusively in short-period binary systems is only 0.09, which is consistent with the findings of this study. 

%####################################################################
\begin{figure}
    \centering
    \resizebox{\hsize}{!}{\includegraphics{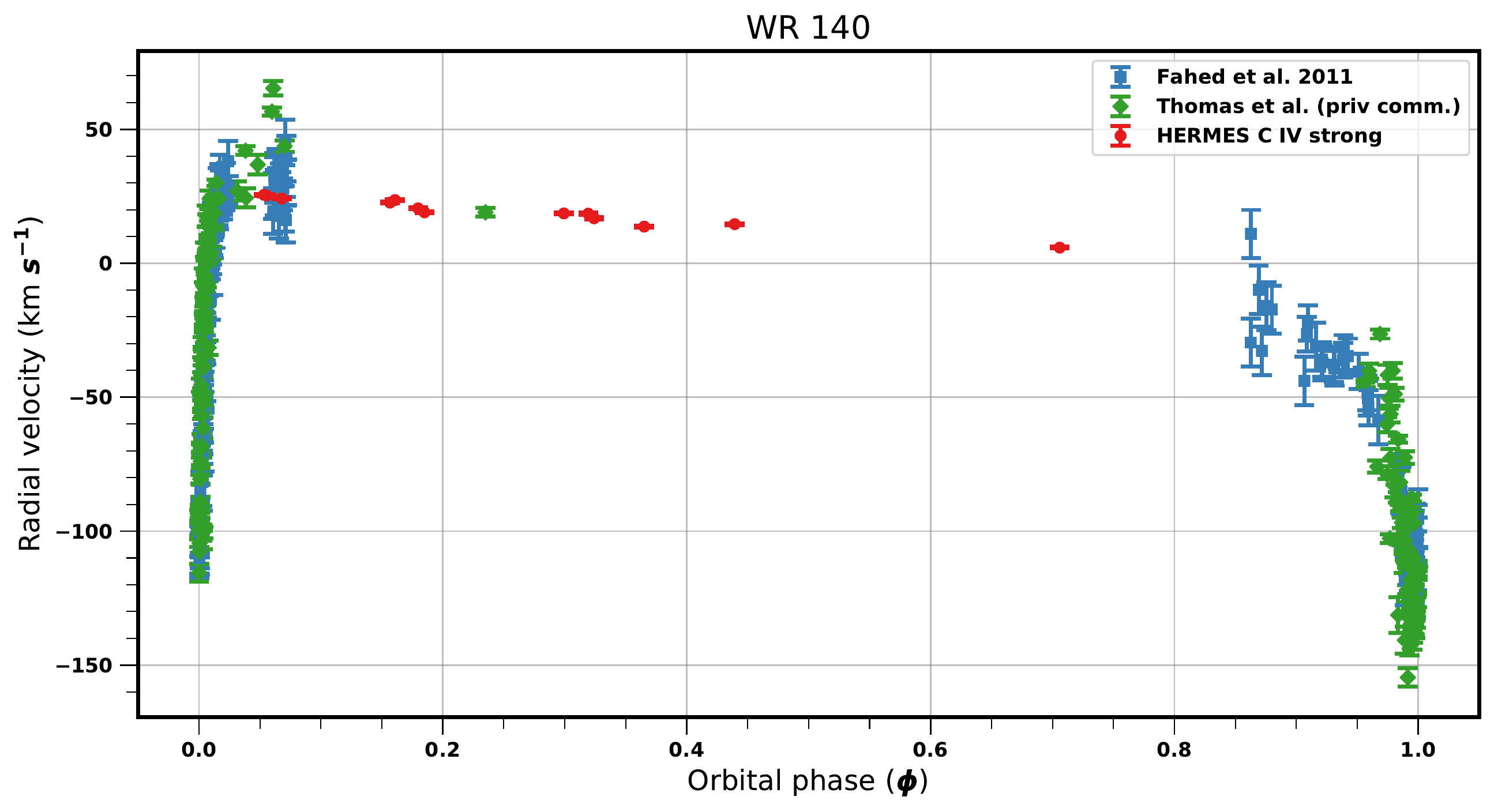}}
    \resizebox{\hsize}{!}{\includegraphics{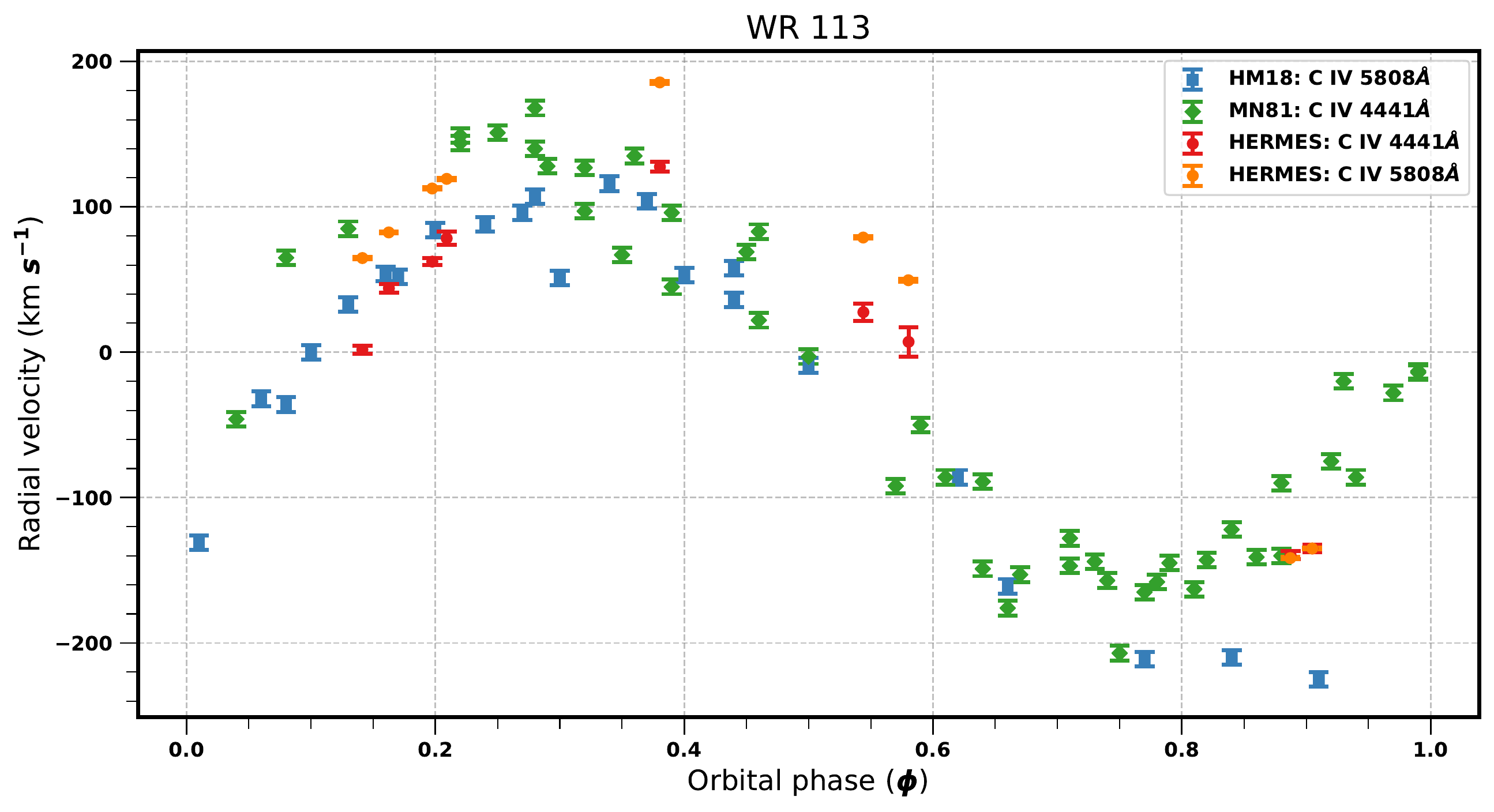}}
    \caption{\textbf{Top:} Phased RV plot for WR 140. Archival RV measurements are shown in blue \citep{2011Fahed} and green (Thomas et al. submitted). The RV measurements from HERMES (red) using C IV strong lines have been shifted by 19~\kms\,in order to convert relative RV measurements to absolute ones. \textbf{Bottom:} Same for WR 113. Archival RV measurements are shown in blue \citep{2018Hill} and green \citep{1981MasseyNiemela}. The RV measurements from HERMES (red and yellow) have been shifted by $-$135~\kms\,in order to convert relative RV measurements to absolute values.}
    \label{fig:known_orbits}
\end{figure}
%####################################################################

\begin{figure}
    \centering
     \resizebox{\hsize}{!}{\includegraphics{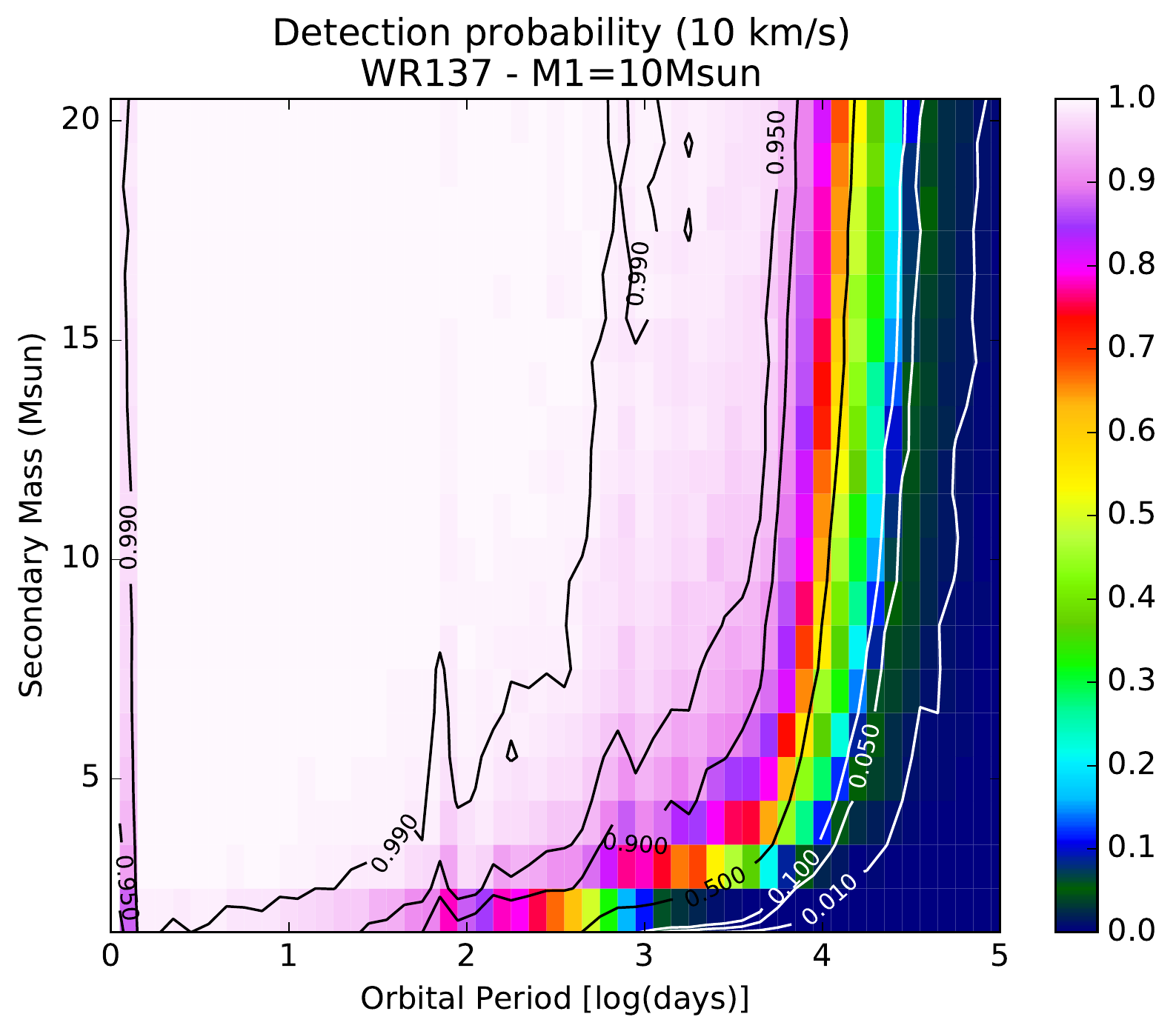}}
    \caption{Detection probability for WR 137 as a function of the orbital period and secondary mass. A representative primary mass 10~M$_{\odot}$ has been adopted for the WR star.}
    \label{f:Pdetect137}
\end{figure}
%####################################################################
\begin{figure}
    \centering
     \resizebox{\hsize}{!}{\includegraphics{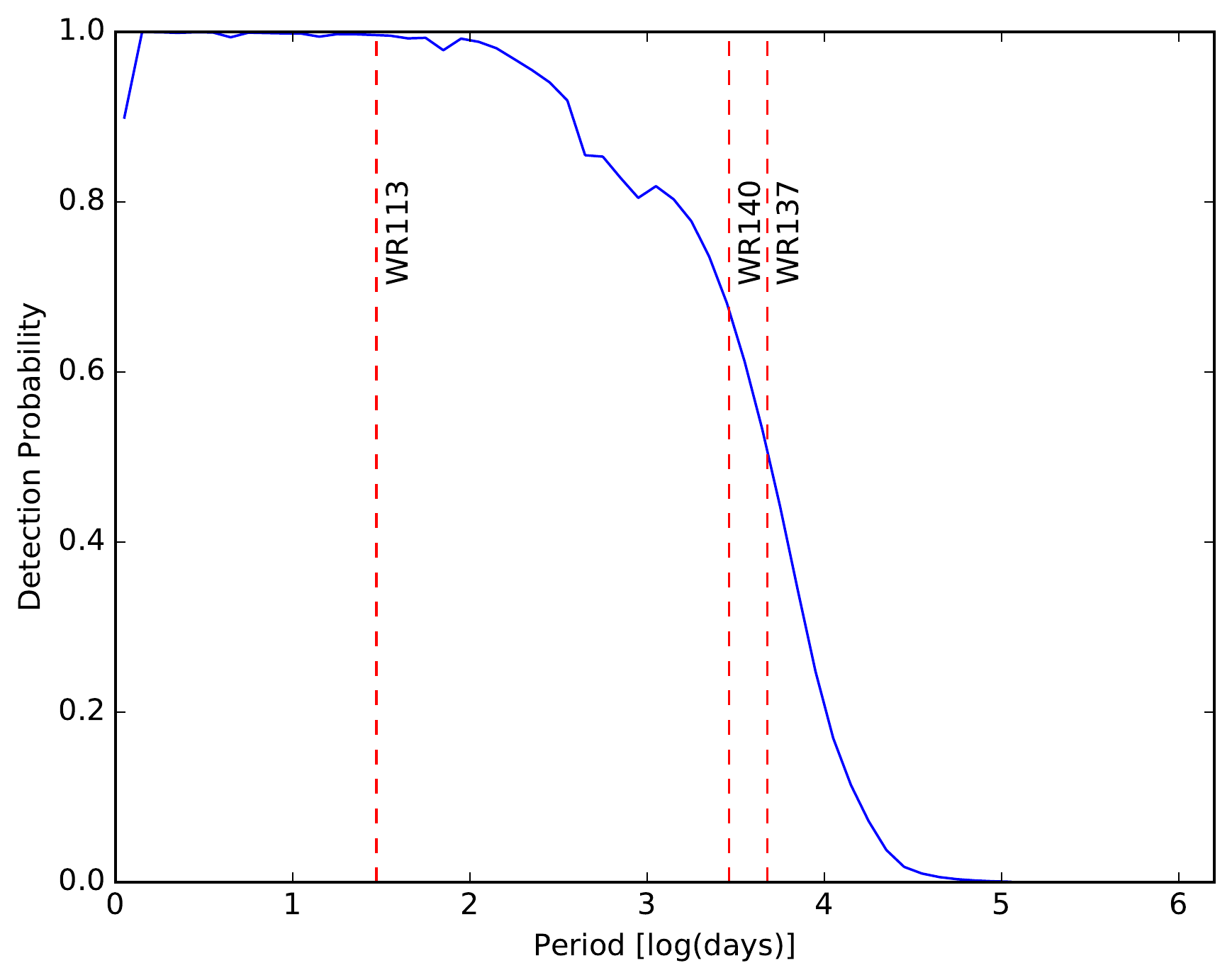}}
    \caption{Average detection probability curve of our campaign as a function of the orbital period and assuming a flat mass-ratio distribution between 0.5 and 2.0. Known orbital periods of WC stars in our sample are indicated by vertical lines.}
    \label{f:Pdetect}
\end{figure}
%####################################################################
\subsection{Instrinsic multiplicity properties}\label{s:intMultiplicityProp}
\subsubsection{Binary detection probability}\label{s:detect}
To estimate the sensitivity domain of our survey, we perform Monte-Carlo (MC) simulations following the method described in \citet{2019Patrick}. In short, for a grid of orbital period and secondary masses ranging from 1 to $10^5$~d and 1 to 20~M$_{\odot}$ respectively, we adopt a flat eccentricity distribution between 0.0 and 0.9, a random orientation of the orbital planes in the three-dimensional space, random times of periastron passage, and RV measurements uncertainties of \SI{1.6}{} km\,s$^{-1}$ that are taken from the single stars in Table \ref{tab:WC_data}. Simulated RV time series of 2500 systems are computed at each $\log P - M_2$ grid cells, with steps of $\Delta \log P =0.1$ and $\Delta M_2=1$~M$_{\odot}$, allowing us to compute the probability that the given orbital configuration would yield a RV signal passing our detection threshold, i.e.\ max$(\Delta \textrm{ RV}) > 10$~km\,s$^{-1}$, given the observational sampling corresponding to each specific object. WC primary masses of 10~M$_{\odot}$ are adopted throughout the simulations \citep[cf.][]{1981Massey}. An example of the resulting detection probability maps is given in Fig.~\ref{f:Pdetect137}. Plots for all other sources in our sample are included in Figs.~\ref{f:Pdetect2} and \ref{f:Pdetect3}. Detection probabilities well over 90\%\ are reached up to periods of $10^{2.5}$ to $10^{3.5}$~d (i.e., 1 to 10 years) depending on the object, except at very low-mass companion masses. Figure ~\ref{f:Pdetect} shows the combined detection probability as a function of the orbital period assuming a flat mass-ratio distribution between 0.5 and 2.0 \citep[e.g., as in][]{2019Langer} and indicates excellent detection rates close to 100\%\ up to periods of 100~d, above 80\%\ up to 1000~d, then dropping to 20\%\ at 10\,000~d.
%---------old P detect--------------------
\subsubsection{Intrinsic multiplicity properties}\label{subs:intMulProp}

The observed binary fraction is a minimum estimate on the number of true WC binaries in our sample. The number of undetected binary systems depends on our time sampling, our RV measurement accuracy, and the intrinsic distribution of orbital parameters. Given our limited sample size, it is beyond the scope of this paper (and probably unrealistic) to perform a self-consistent derivation of the distributions of orbital parameters, and of the binary fraction  \citep[as in, e.g., ][]{2012Sana, 2013Sana}. Nevertheless, it is still possible to obtain useful information on the multiplicity properties of the parent population. In this context, we adopt two different and complementary approaches, both using MC simulations and the same underlying assumptions as in Sect.~\ref{s:detect}. 

% The first step focuses on reconstructing the underlying period distribution of our sample. For this purpose, we compute for each object $k$, the  probability $p_k(P,q)$ of each combination of orbital period $P$ and mass-ratio $q$ to reproduce the measured $\max(\Delta \textrm{ RV})$-signal within the 95\%\ confidence interval ($\pm3.2$~\kms) of the observed value. After normalising the probability to unity over a given $P-q$ grid, we sum up and collapse along the mass-ratio axis to obtain the likely period distribution (Fig.~\ref{f:Pdist}). We perform this exercise with the full sample in two different scenarios: by implicitly assuming that all the objects in our sample are binaries, and that only seven out of twelve objects pass our binary detection threshold. In either case, the emergent picture is the same (Fig.~\ref{f:Pdist}): a broad peak at long periods in the range of $10^3$ to $10^4$~d and a low-level plateau at shorter period that amount for about 10\%\ of the number of binaries in the peak.

As a first step, we attempt to constrain the true binary fraction among our sample of WC stars. A simple approach would be to correct the number of detected binaries using a detection probability factor that depends on the orbital period of the detected binaries. This probability can be readily read from Fig.~\ref{f:Pdetect}. Unfortunately, only three of the objects above our binary threshold have known orbital periods (see Table~\ref{tab:WC_data}). Lacking further information, we conservatively adopt the detection probability corresponding to period of $10^{3.5}$ d (approximately \ 10 years) for the remaining objects (shorter periods can be excluded on the basis of the previous step, as they would have yield a RV-amplitude larger than 30~\kms\ in over 90\%\ of the cases). Under these hypotheses, the obtained correction factor is found to be of the order of 3/2, suggesting an intrinsic binary fraction of about 0.9 for the objects in our sample in the considered parameter space (P up to $10^5$~d and mass-ratios up to 3.0).

To test these intriguing results, we set up a last experiment where we test different period distributions and intrinsic binary fractions with the same assumptions on $q$, $i$, and $e$ as in Sect.\ref{s:detect}. We reject combinations of orbital configurations and binary fractions that cannot reproduce a simple observational fact: six objects are observed with values of $\Delta$RV between 10 and 30 \kms\,within 1-$\sigma$ (Fig.~\ref{fig:bin_frac})\footnote{Similar results were obtained when trying to reproduce the number of stars with RV amplitude between 15 and 30~km\,s$^{-1}$.}. We constructed two experiments based on: 
\begin{enumerate}
    \item[-] a uniform $\log P$  distribution with variable minimum and maximum boundaries, and
    \item[-] a normal $\log P$  distribution with variable central period and $1\sigma$-dispersion.
\end{enumerate} 
For these experiments, we removed WR~113 from the sample as its significantly different orbital period would require more complicated functional forms than those adopted here. In that case, we also request that no system with an RV variation amplitude above 30~\kms\ be found.

In both cases, parent distributions with intrinsic binary fractions larger than 90\%\ performed significantly better than those with lower binary fractions. This is in line with our above estimate based on the detection probability. 
% Both the uniform and the normal distributions suggest a rather narrow peak centered around $10^3$ to $10^{4}$~d. 

\subsubsection{Intrinsic binary fraction of the parent WC population}
Overall, the results of these MC experiments suggest that WC stars in our sample have a large binary fraction and are dominated by long-period systems. The present sample was selected based on declination and magnitude limits and should not contain any specific selection bias. Therefore, it should remain representative of the entire WC population. Under these considerations, a test-of-hypothesis at the 10\%-significance level allows us to reject parent binary fractions smaller than 0.72\footnote{Specifically, binomial chances to draw 11 or more binaries in a sample of 12 (corresponding the retrieved intrinsic binary fraction in the sample of $>0.9$) drops below 10\% for parent binary fraction smaller than 0.72.}. This suggests that, with a 90\% confidence limit, the true binary fraction of the carbon-sequence WR population in the Milky Way lays in the range 0.72 to 1.00. In contrast to the results in Sect. \ref{subs:intMulProp}, the latter value does account for the sample size.

\section{Evolutionary implications and conclusions}\label{sect:conclusions}
%-------------------------------------------------------------------
With a small, magnitude-limited sample of northern Galactic WC stars, we have demonstrated that using cross-correlation with high-resolution spectra allows us to derive accurate RV measurements of WR stars, and to even quantify the wind-variability. With a sample of 12 stars spread evenly over different spectral-types, we find an observed binary fraction of 0.58. Using MC simulations, we attempt to gain information about the parent WC population and conclude that the fraction of Galactic WC stars that are part of a binary system is at least 0.72, significantly larger than the literature value of ${\sim}$0.4. 

% Pertaining future validation of the multiplicity properties, we briefly discuss some of the evolutionary implications of our results {\it assuming} they are validated by further observations. 
Our observations clearly show that there is a deficiency in the number of short-period WC binaries in the Galaxy. Furthermore, our MC simulations indicate that Galactic WC stars primarily exist in long-period binaries (P $>$ 100~d). Comparing these results to the observed population of main-sequence O-type binaries can help us put constraints on the viable evolutionary channels for WC stars and on the forward evolution of massive binaries. We briefly discuss these evolutionary channels below.

O-type stars are generally expected to evolve through the modified Conti scenario \citep[O$\rightarrow$LBV$\rightarrow$WN$\rightarrow$WC$\rightarrow$WO,][]{2007Crowther}, although the dominant mechanism leading to the formation of WR stars is still debated. Keeping this in mind, we briefly look at the outcome of possible evolutionary channels for short-period O-type binaries. These binaries can interact through Roche Lobe overflow (RLOF), common-envelope evolution, or even through mergers. We explore possible mechanisms that can explain the formation of long-period WC binaries when starting from short-period O-type binaries. 

A first viable scenario that produces long-period binaries is the RLOF channel. Depending on the initial orbital configuration, the primary may undergo Case-A (on the main sequence) or early Case-B (post main sequence) mass transfer. The companion will, thus, strip the primary of its hydrogen envelope, leading to the formation of a WN star. Except in the case of a low-mass companion, this results in a mass-ratio inversion, irrespective of whether the mass transfer is conservative or non-conservative. This then leads to the widening of the orbit. Further loss of mass through winds during the entire WN and the early WC phase adds to the widening of the orbit. 

The simulation study by \citet{2019Langer} explores the RLOF channel, albeit with the aim of constraining the period distribution of OB+BH binary systems (top panel of their Figure 6). Given the fact that WC stars are thought to be the direct progenitors of BHs and assuming that the orbit does not undergo drastic change during the short-lived WC phase, this distribution should be valid for WC binary systems. The distribution has a small peak at short periods (log~P $\sim$ 0.7) and a larger peak at long periods (log~P $\sim$ 2.2). It is important to note that the simulations in \citet{2019Langer} are performed with MESA binary evolution tracks at lower metallicity, for the Large Magellanic Cloud. At Galactic metallicity, the corresponding evolutionary tracks will have higher mass-loss rates and hence, shift the distribution towards longer periods. Coupled with the fact that the theoretical distribution of \citet{2019Langer} can contain BHs formed from WN stars, it is plausible that our observations sample the tail of the period distribution. 

Another evolutionary scenario focuses on common-envelope evolution and the possibility of a merger. Common-envelope evolution leads to the shrinking of the orbit, resulting in a very short-period binary system that may merge. In the case of a merger, the binary system ceases to exist and may result in an apparently single WC star. This channel may partly be responsible for the formation of the apparently-single WC stars in our sample. However, it is also possible that these stars formed as single stars. 

A final channel to consider that produces long-period binaries invokes the existence of a hierarchical triple system \citep[e.g.][]{2014Sana,2016MaizApellaniz}. If the outer component can evolve into a WR due to wind-stripping then it can be observed as a component in a wide binary system. Alternatively, the merger of the inner binary followed by strong mass loss can also lead to the formation of a WR star, leading to multiple possibilities. These scenarios invoke the formation of a WN star by wind-stripping, in contrast to companion-stripping. Lastly, the expansion of the inner system might lead to dynamical interaction with the tertiary companion, potentially resulting in the expulsion of one of the stars in the system.

In conclusion, these results exhibit that there is a dire need to increase the sample by expanding RV surveys to the southern sky and to fainter magnitudes. This will enable us to derive the binary fraction and the intrinsic distributions of the orbital parameters of massive stars at the end stages of their lives in a self-consistent manner. This combined with what is known for main sequence O-type stars, is the key to advancing our knowledge of massive binary evolution. 

\begin{acknowledgements}
KD, TS and HS acknowledge support from the European Research Council (ERC) innovation programme under the European Union’s DLV-772225-MULTIPLES Horizon 2020 research and innovation programme. PM acknowledges support from the FWO junior postdoctoral fellowship No. 12ZY520N.
\end{acknowledgements}

%-------------------------------------------------------------------

\bibliographystyle{aa} % style aa.bst
\bibliography{references} % your references references.bib

\appendix
\section{Comments on specific objects}\label{apdx:comments}
%####################################################################
% \subsection{WR 4}

\textbf{WR 4:} According to the seventh catalogue of WR stars \citep{2001vanderHucht}, WR 4 is classified as a WC5+?, an SB1 with no observable line dilution or absorption effects. Photometric variability with a period of 4.3 days \citep{1986MoffatShara} was determined. \citet{1989RustamovCherepashchuk} found both a photometric and spectroscopic period of 2.4096 days with a peak-to-peak amplitude of ${\sim}$60\kms, but we can safely reject the presence of such large amplitude RV variation in our data set (Table \ref{tab:WC_data}) questioning the fact that the 2.4~d period traces  an orbital motion. We therefore do not include WR 4 in the list of short-period binaries in our literature study (Sect. \ref{subsec:GWC_bin}).

In a study of co-rotating interaction regions (CIRs), \citet{2009StLouis} classified WR 4 as `Not Variable' in the weak C IV \SI{5016}{\angstrom} line. \citet{2011CheneStLouis} further elaborated that the maximum line-profile variability was 1.2\% of the level of flux in the line, hence making it ineligible to be a CIR candidate. In this work, we obtained 22 epochs over a time coverage of 890 days, out of which about 12 were observed within two nights, attempting a probe on shorter periods. Although we do not report a short period, the adopted choice of our threshold allows us to classify it as a binary (Table \ref{tab:WC_data}). Further monitoring of this system is required to determine the orbital period of this system.

% \subsection{WR 5}

\textbf{WR 5:} \citet{2001vanderHucht} reports WR 5 as a single star with spectral type of WC6. No emission was found in the X-rays \citep{Skinner2006,2009GudelNaze}. In this work, WR 5 is found to have a $\Delta$ RV of \SI{4.9}{\km\per\s} over 718 days with a $\sigma_{RV}$ of \SI{1.7}{\km\per\s} for RVs measured using the full spectrum. It does not pass our binary detection criterion and is, therefore, classified as presumably single.

% \subsection{WR 111}

\textbf{WR 111:} WR 111 is a presumably single WC5 star \citep{1999aNiemela,2001vanderHucht}. It has been studied with spectral analyses and hydrodynamical studies \citep{2002Grafener,2005GrafenerHamann,2012Sander}, and was used to show for the first time that WC winds can be explained through radiation-driven mass loss. In this work, WR 111 shows a peak-to-peak variability of \SI{4.5}{\km\per\s} over 742 days with a $\sigma_{RV}$ of \SI{1.5}{\km\per\s}. It does not pass our binary detection criterion and is therefore classified as a presumably single star.

% \subsection{WR 113}

\textbf{WR 113:} WR 113 is a WC8d+O8-9 binary system first identified by \citet{1981MasseyNiemela} with a period of about 29.707 days in a circular orbit. A debate about the eccentricity arose when \citet{1996Niemela} derived similar orbital parameters with a significant eccentricity ($e=0.19$). The matter was resolved by \citet{2012David-Uraz}, who showed that the two were indistinguishable solutions and adopted a circular orbit and was later confirmed by \citet{2018Hill} with a period of 29.700 $\pm$ 0.001 d. In this work, WR 113 has the largest $\Delta$ RV and $\sigma_{RV}$ of 282.6 and \SI{93.9}{\km\per\s} respectively, passing our binary detection criterion. The RV measurements phased with the orbital period are shown in Fig. \ref{fig:known_orbits}. 

% \subsection{WR 117}

\textbf{WR 117:} According to the GCWR, WR 117 is classified as a WC9d, which is a dust producing star. \citet{2012Sander} find it to be a bit too hotter than most WC9 counterparts, even suggesting it might be in the WC8 parameter regime. \citet{2005Williams} did not find any evidence for an OB-type companion, and hypothesize that WR 117 might be an evolved WC9 star. A large scatter in photometric magnitudes was observed by \citet{2014Williams} with possible 0.5 mag eclipses due to dust. In this work, WR 117 shows a $\Delta$ RV of \SI{4.9}{\km\per\s} and a $\sigma_{RV}$ of \SI{1.9}{\km\per\s} over a coverage of 488 days and is hence classified as a presumably single star.

% \subsection{WR 119}
\textbf{WR 119:} WR 119 is classified as a single WC9d star according to the GCWR. \citet{2009Fahed} studied it using $V$ and $I$ band photometry, finding a larger than normal variability in these bands. \citet{2017Desforges} performed a temporal variance spectrum analysis over timescales of days to weeks and found significant variability in the  C III  \SI{5696}{\angstrom} line. In this work, we find values of $\Delta$ RV and $\sigma_{RV}$ of 11.0 and \SI{3.2}{\km\per\s} respectively, resulting in WR 119 being classified as a binary. 
% The RV plot over time can be seen in Appendix 3? \textbf{(tbm)}. Whether this variability is intrinsic or due to orbital motion is a question that will require more observations to answer.

% \subsection{WR 135}

\textbf{WR 135:} WR 135 is an apparently single WC8 star (GCWR). Previous RV studies have failed to determine any variability \citep{1999aNiemela}, nor even photometric variability \citep{1986MoffatShara}. In this work, we find a $\Delta$ RV of \SI{4.7}{\km\per\s} and $\sigma_{RV}$ of \SI{1.6}{\km\per\s} over a baseline of 2660 days, classifying it as a presumably single star.

% \subsection{WR 137}
\textbf{WR 137:} WR 137 is a well-studied, long-period binary system with a late-type WC star and a late-type O star (WC7pd + O9 V; p: peculiar, d: dust producing). \citet{1999MarchenkoDust} reported dust production in 1997 September - 1998 May and \citet{2001WilliamsDust} derived a period of 13.05 $\pm$ 0.15 years by tracking the periodic dust formation. This period was verified by the orbital analysis using RVs by \citet{2005Lefevre}, who found the maximum emission from dust to occur just after the periastron passage. The orbit was recently resolved using interferometry with CHARA \citep{2016Richardson}, reiterating the importance of long term RV monitoring campaigns in identifying WR binaries. The broad emission lines of the WR star are superimposed by a narrow emission and an absorption component, making the companion star a prime candidate to be an Oe star. The RVs measured in this campaign are in agreement with those obtained from the literature (Fig. \ref{fig:WR137_shortcadence}).

% \subsection{WR 140}

\textbf{WR 140:} WR 140 is classified as a (WC7pd + O5.5fc) system with a well established spectroscopic \citep{2011Fahed} and visual \citep[interferometric;][]{2011Monnier} orbit (P=7.938 years, e=0.8996). These parameters were derived by combining archival data along with measurements during the 2009 periastron passage. An improved derivation of the masses and orbital parameters using spectroscopic and interferometric data from the 2016 December periastron passage makes it one of the most accurately constrained masses for a WR binary system to date (Thomas et al. submitted). In this work, WR 140 has a $\Delta$RV of 21.6\kms\, and a $\sigma_{RV}$ of 6.5\kms, which passes our binary detection criterion. The RV measurements along with archival data are shown in Fig. \ref{fig:known_orbits}.

% \subsection{WR 143}

\textbf{WR 143:} WR 143 is classified as a binary system \citep[WC4 + Be;][]{2006VarricattAshok}, who detected the presence of hydrogen emission lines and He I lines in the infrared spectrum. In the optical spectra, a hint of this is seen in H$\alpha$, with features similar to those exhibited by WR 137. We detect significant RV variability over our 76~d baseline and thus classify WR 143 as a binary ($\Delta$ RV = \SI{11.2}{\km\per\s}, $\sigma_{RV}$ = \SI{4.2}{\km\per\s}).

% \subsection{WR 146}

\textbf{WR 146:} WR 146 is a visual binary with a spectral classification of WC5 + O8. The components show non-thermal radio emission \citep{1996Dougherty} and were resolved with HST images \citep{1998Niemela}. The computed mass loss rate for WR 146 is much higher than prototypical WC5 stars, prompting a debate on the multiplicity of the O-star companion \citep{2000Dougherty}. An X-ray study by \citet{2017Zhekov} shows a discrepancy in the mass-loss rate by a factor of 8 to 10. Given the angular separation, \citet{1998Niemela} suggest binary periods of hundreds of years. However, our 847 day baseline with values for $\Delta$ RV and $\sigma_{RV}$ of 18.9 and \SI{5.9}{\km\per\s} suggest periods of the order of 3000 d (Fig. \ref{f:Pdetect2}) for a massive companion. Further monitoring is required to constrain the orbit. 

% \subsection{WR 154}

\textbf{WR 154:} WR 154 is classified as a single WC6 star (GCWR). In this work, we do not detect significant RV variation. However, given the short baseline (24 days), we cannot rule out the possibility of it being a long period binary as the detection probability of our campaign drops significantly for periods above 100~d (Fig. \ref{f:Pdetect2}).
%-------------------------------------------------------------------

\section{Relative RV measurements}\label{s:tables_RV}
Relative RVs for the objects in our sample, measured with the full spectrum as a mask. The reference epoch is marked with a `*'. The Barycentric Julian Date (BJD) is given as the middle of the exposure. The average S/N is given in Table\,\ref{tab:epochs}.
\begin{table}[h!]
    \centering
    \caption{Journal of the HERMES observations for WR 4}
    \begin{tabular}{ccc} \hline \hline
        BJD $-$ 2450000 (d) & Relative RV (\kms) & $\sigma_p$ (\kms) \\ \hline
        7961.7217 & 1.39 & 0.12 \\ 
        8088.5764$^*$ & 0.00 & 0.13 \\ 
        8103.4795 & $-$3.18 & 0.10 \\ 
        8712.7295 & 2.32 & 0.10 \\ 
        8719.6991 & 0.95 & 0.09 \\ 
        8740.6337 & 0.08 & 0.11 \\ 
        8779.4835 & 0.52 & 0.09 \\ 
        8779.4945 & 0.39 & 0.09 \\ 
        8779.5055 & 0.57 & 0.09 \\ 
        8779.5165 & 1.41 & 0.09 \\ 
        8779.5275 & 0.86 & 0.09 \\ 
        8784.6062 & 0.99 & 0.17 \\ 
        8784.6276 & 1.62 & 0.16 \\ 
        8784.6490 & 1.71 & 0.12 \\ 
        8784.6705 & 2.20 & 0.12 \\ 
        8784.6919 & 2.51 & 0.12 \\ 
        8796.5122 & 3.44 & 0.11 \\ 
        8798.5655 & 0.02 & 0.12 \\ 
        8799.5284 & $-$6.57 & 0.62 \\ 
        8804.6853 & $-$7.21 & 0.60 \\ 
        8821.4885 & 3.53 & 0.12 \\ 
        8852.3716 & 17.67 & 0.11 \\ \hline
    \end{tabular}
    %\label{tab:my_label}
\end{table}

\begin{table}[h!]
    \centering
    \caption{Journal of the HERMES observations for WR 5}
    \begin{tabular}{ccc} \hline \hline
        BJD $-$ 2450000 (d) & Relative RV (\kms) & $\sigma_p$ (\kms) \\ \hline
        8086.5338 & $-$2.36 & 0.11 \\
        8103.4889 & $-$4.88 & 0.09 \\
        8721.7268 & $-$1.05 & 0.08 \\
        8742.6913 & $-$0.75 & 0.11 \\
        8756.7129$^*$ & 0.00 & 0.08 \\
        8789.6356 & $-$4.21 & 0.28 \\ 
        8804.7296 & $-$1.57 & 0.09 \\
        \hline
    \end{tabular}
    %\label{tab:my_label}
\end{table}

\begin{table}[h!]
    \centering
    \caption{Journal of the HERMES observations for WR 111}
    \begin{tabular}{ccc} \hline \hline
        BJD $-$ 2450000 (d) & Relative RV (\kms) & $\sigma_p$ (\kms) \\ \hline
        7881.5925 & 2.59 & 0.06 \\ 
        7900.5403 & $-$0.63 & 0.07 \\ 
        8249.7094$^*$ & 0.00 & 0.04 \\ 
        8276.6582 & 2.81 & 0.05 \\ 
        8276.6760 & 2.27 & 0.06 \\ 
        8598.6858 & 4.23 & 0.05 \\ 
        8621.6290 & 0.14 & 0.11 \\ 
        8623.5349 & 2.93 & 0.05 \\ 
        \hline
    \end{tabular}
    %\label{tab:my_label}
\end{table}

\begin{table}[h!]
    \centering
    \caption{Journal of the HERMES observations for WR 113}
    \begin{tabular}{ccc} \hline \hline
        BJD $-$ 2450000 (d) & Relative RV (\kms) & $\sigma_p$ (\kms) \\ \hline
        904.7122 & 265.36 & 0.46 \\ 
        7909.5689 & 117.86 & 1.10 \\
        7910.6482 & 95.97 & 0.74 \\
        8254.6411 & 249.11 & 0.23 \\
        8255.6767 & 259.04 & 0.18 \\
        8276.6878$^*$ & 0.00 & 0.27 \\
        8580.7046 & 240.54 & 0.34 \\
        8582.7191 & 262.29 & 0.26 \\
        8632.5666 & $-$17.27 & 0.29 \\
        \hline
    \end{tabular}
    %\label{tab:my_label}
\end{table}

\begin{table}[h!]
    \centering
    \caption{Journal of the HERMES observations for WR 117}
    \begin{tabular}{ccc} \hline \hline
        BJD $-$ 2450000 (d) & Relative RV (\kms) & $\sigma_p$ (\kms) \\ \hline
        8254.7002 & 2.45 & 0.44 \\ 
        8255.7058$^*$ & 0.00 & 0.33 \\ 
        8701.3976 & 4.52 & 0.36 \\ 
        8742.4328 & 1.86 & 0.39 \\ 
        \hline
    \end{tabular}
    %\label{tab:my_label}
\end{table}

\begin{table}[h!]
    \centering
    \caption{Journal of the HERMES observations for WR 119}
    \begin{tabular}{ccc} \hline \hline
        BJD $-$ 2450000 (d) & Relative RV (\kms) & $\sigma_p$ (\kms) \\ \hline
        7906.7165 & 10.97 & 0.25 \\ 
        7908.6325 & 6.35 & 0.26 \\ 
        7911.5561 & 4.93 & 0.26 \\ 
        8282.6811$^*$ & 0.00 & 0.12 \\ 
        8283.5768 & 0.74 & 0.19 \\ 
        8723.4382 & 7.27 & 0.22 \\ 
        8727.4160 & 5.99 & 0.34 \\ 
        \hline
    \end{tabular}
    %\label{tab:my_label}
\end{table}s

\begin{table}[h!]
    \centering
    \caption{Journal of the HERMES observations for WR 135}
    \begin{tabular}{ccc} \hline \hline
        BJD $-$ 2450000 (d) & Relative RV (\kms) & $\sigma_p$ (\kms) \\ \hline
        6119.7192 & $-$2.94 & 0.06 \\
        6119.7247 & $-$2.92 & 0.06 \\
        6119.7301 & $-$2.88 & 0.06 \\
        7309.4567$^*$ & 0.00 & 0.04 \\
        7310.3944 & $-$1.73 & 0.05 \\
        7900.7233 & $-$1.14 & 0.04 \\
        7935.6002 & $-$1.54 & 0.05 \\
        8197.7621 & 1.30 & 0.05 \\
        8206.7455 & $-$1.81 & 0.05 \\
        8267.5912 & $-$1.45 & 0.04 \\
        8276.7069 & $-$1.72 & 0.04 \\
        8276.7155 & $-$1.79 & 0.04 \\
        8629.7216 & $-$2.82 & 0.04 \\
        8663.7244 & $-$0.31 & 0.05 \\
        8719.5508 & 1.60 & 0.09 \\
        8779.4288 & $-$3.12 & 0.06 \\
        \hline
    \end{tabular}
    %\label{tab:my_label}
\end{table}

\begin{table}[]
    \centering
    \caption{Journal of the HERMES observations for WR 137}
    \begin{tabular}{ccc} \hline \hline
        BJD $-$ 2450000 (d) & Relative RV (\kms) & $\sigma_p$ (\kms) \\ \hline
        6125.6634 & 17.28 & 0.17 \\
        6889.6039 & $-$1.07 & 0.13 \\
        7324.4165 & $-$5.66 & 0.12 \\
        7902.7153 & $-$4.62 & 0.16 \\
        7938.6925 & $-$7.84 & 0.16 \\
        7951.5912 & $-$6.72 & 0.18 \\
        8195.7661 & $-$4.58 & 0.17 \\
        8206.7607 & $-$3.57 & 0.16 \\
        8265.6372 & $-$5.00 & 0.16 \\
        8629.7139 & $-$0.47 & 0.13 \\
        8664.7263 & 2.19 & 0.13 \\
        8705.5661 & 0.02 & 0.10 \\
        8706.6067 & 0.68 & 0.09 \\
        8707.5508 & 0.32 & 0.12 \\
        8708.4941 & $-$0.37 & 0.08 \\
        8709.4256 & 0.24 & 0.10 \\
        8710.5560 & 0.76 & 0.10 \\
        8712.5473 & 0.09 & 0.10 \\
        8713.5166 & 0.35 & 0.11 \\
        8714.4978 & $-$0.27 & 0.09 \\
        8715.5106 & 0.40 & 0.12 \\
        8716.5440$^*$ & 0.00 & 0.12 \\
        8717.5925 & 0.22 & 0.16 \\
        8718.4792 & $-$2.96 & 0.83 \\
        8719.4324 & $-$3.19 & 0.82 \\
        8720.4046 & 1.66 & 0.18 \\
        8721.5327 & $-$1.56 & 0.10 \\
        8779.4381 & 0.53 & 0.15 \\
        \hline
    \end{tabular}
    %\label{tab:my_label}
\end{table}

\begin{table}[h!]
    \centering
    \caption{Journal of the HERMES observations for WR 140}
    \begin{tabular}{ccc} \hline \hline
        BJD $-$ 2450000 (d) & Relative RV (\kms) & $\sigma_p$ (\kms) \\ \hline
        6118.7347 & $-$4.89 & 0.17 \\
        6890.5970 & $-$14.41 & 0.22 \\
        7896.7053 & 6.86 & 0.13 \\
        7902.6990 & 7.19 & 0.12 \\
        7938.6987 & 5.63 & 0.13 \\
        8195.7710 & 3.58 & 0.13 \\
        8207.7645 & 3.94 & 0.12 \\
        8262.6720 & 1.16 & 0.12 \\
        8277.7089$^*$ & 0.00 & 0.10 \\
        8608.7327 & $-$1.83 & 0.16 \\
        8666.6980 & $-$3.24 & 0.16 \\
        8680.5209 & $-$4.25 & 0.20 \\
        8799.4178 & $-$8.28 & 0.23 \\
    \end{tabular}
    %\label{tab:my_label}
\end{table}

\begin{table}[h!]
    \centering
    \caption{Journal of the HERMES observations for WR 143}
    \begin{tabular}{ccc} \hline \hline
        BJD $-$ 2450000 (d) & Relative RV (\kms) & $\sigma_p$ (\kms) \\ \hline
        8723.5075 & 6.39 & 0.37 \\
        8727.5533 & 6.74 & 0.41 \\
        8759.5106$^*$ & 0.00 & 0.43 \\
        8775.3291 & $-$1.09 & 0.59 \\
        8775.3505 & $-$1.15 & 0.62 \\
        8799.4685 & $-$4.49 & 1.53 \\ 
        \hline
    \end{tabular}
    %\label{tab:my_label}
\end{table}

\begin{table}[h!]
    \centering
    \caption{Journal of the HERMES observations for WR 146}
    \begin{tabular}{ccc} \hline \hline
        BJD $-$ 2450000 (d) & Relative RV (\kms) & $\sigma_p$ (\kms) \\ \hline
        7912.6682 & $-$3.55 & 0.67 \\
        7938.7179 & $-$11.18 & 1.41 \\
        7948.7207 & $-$8.07 & 1.64 \\
        8201.7401 & $-$4.40 & 0.35 \\
        8310.6838 & 7.67 & 0.75 \\
        8740.4677 & $-$1.98 & 0.40 \\
        8741.5211 & $-$1.94 & 0.40 \\
        8759.3867$^*$ & 0.00 & 0.38 \\
        \hline
    \end{tabular}
    %\label{tab:my_label}
\end{table}

\begin{table}[h!]
    \centering
    \caption{Journal of the HERMES observations for WR 154}
    \begin{tabular}{ccc} \hline \hline
        BJD $-$ 2450000 (d) & Relative RV (\kms) & $\sigma_p$ (\kms) \\ \hline
        8762.6141 & $-$2.13 & 0.10 \\
        8763.5412 & $-$3.41 & 0.11 \\
        8775.5248 & $-$1.75 & 0.11 \\
        8786.5511$^*$ & 0.00 & 0.13 \\
        \hline
    \end{tabular}
    \label{tab:my_label}
\end{table}
%-------------------------------------------------------------------
\newpage
% \section*{Plots for the detection probability for individual objects}
\begin{figure*}[h!]
    \centering
    \includegraphics[width=8cm]{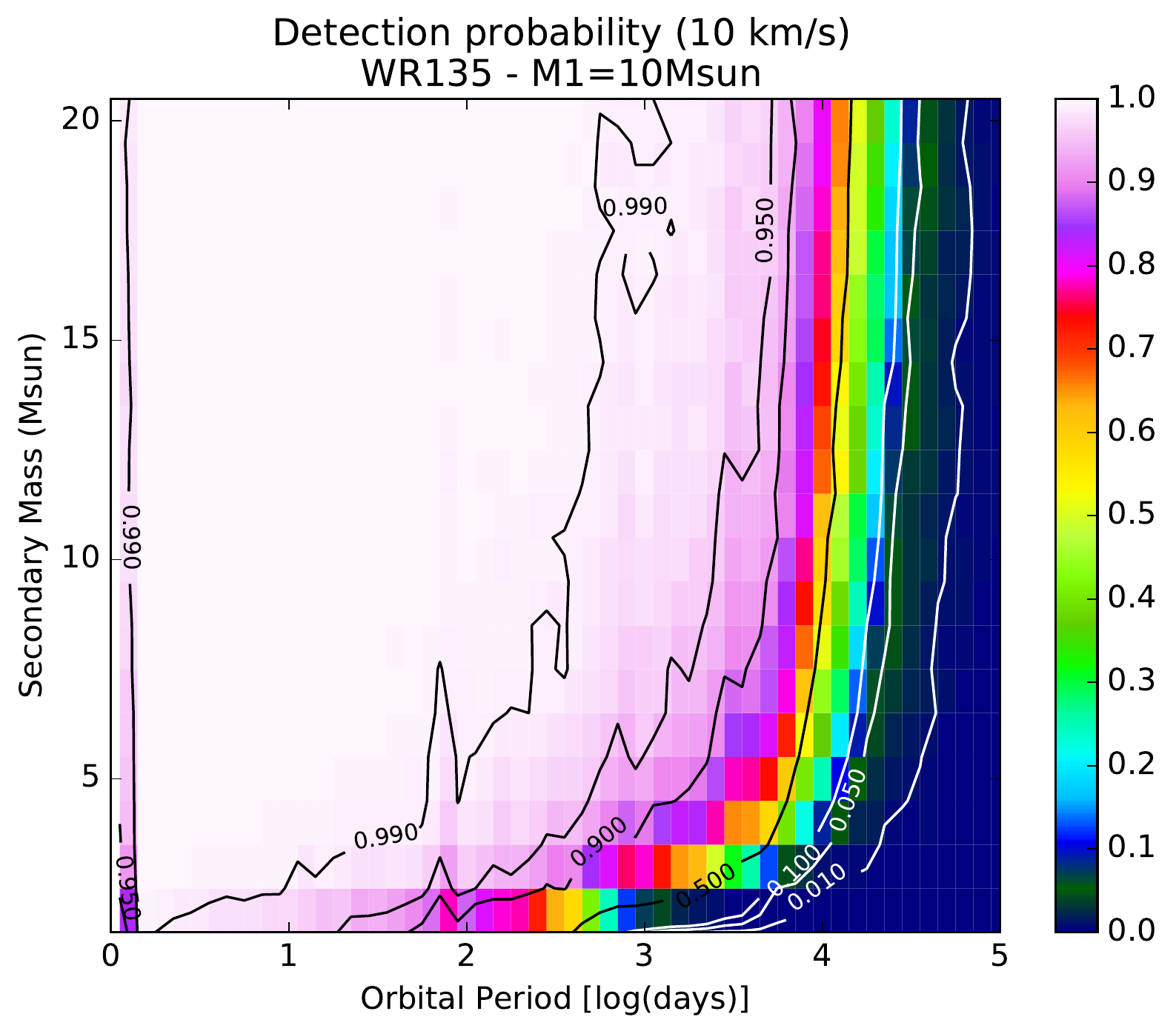}
    \includegraphics[width=8cm]{Figures/137PM2_thres10_MAR31.pdf}
    \includegraphics[width=8cm]{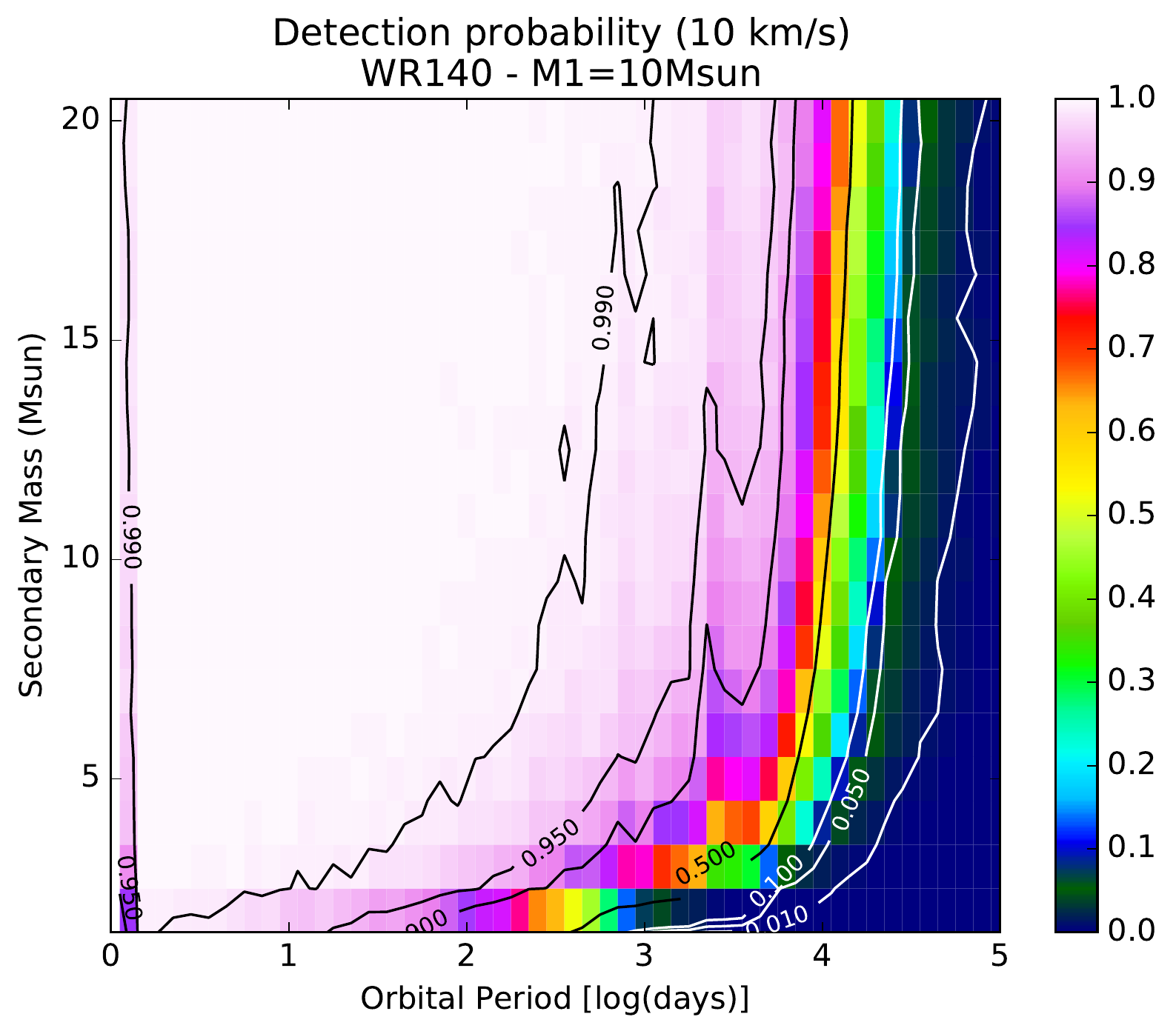}
    \includegraphics[width=8cm]{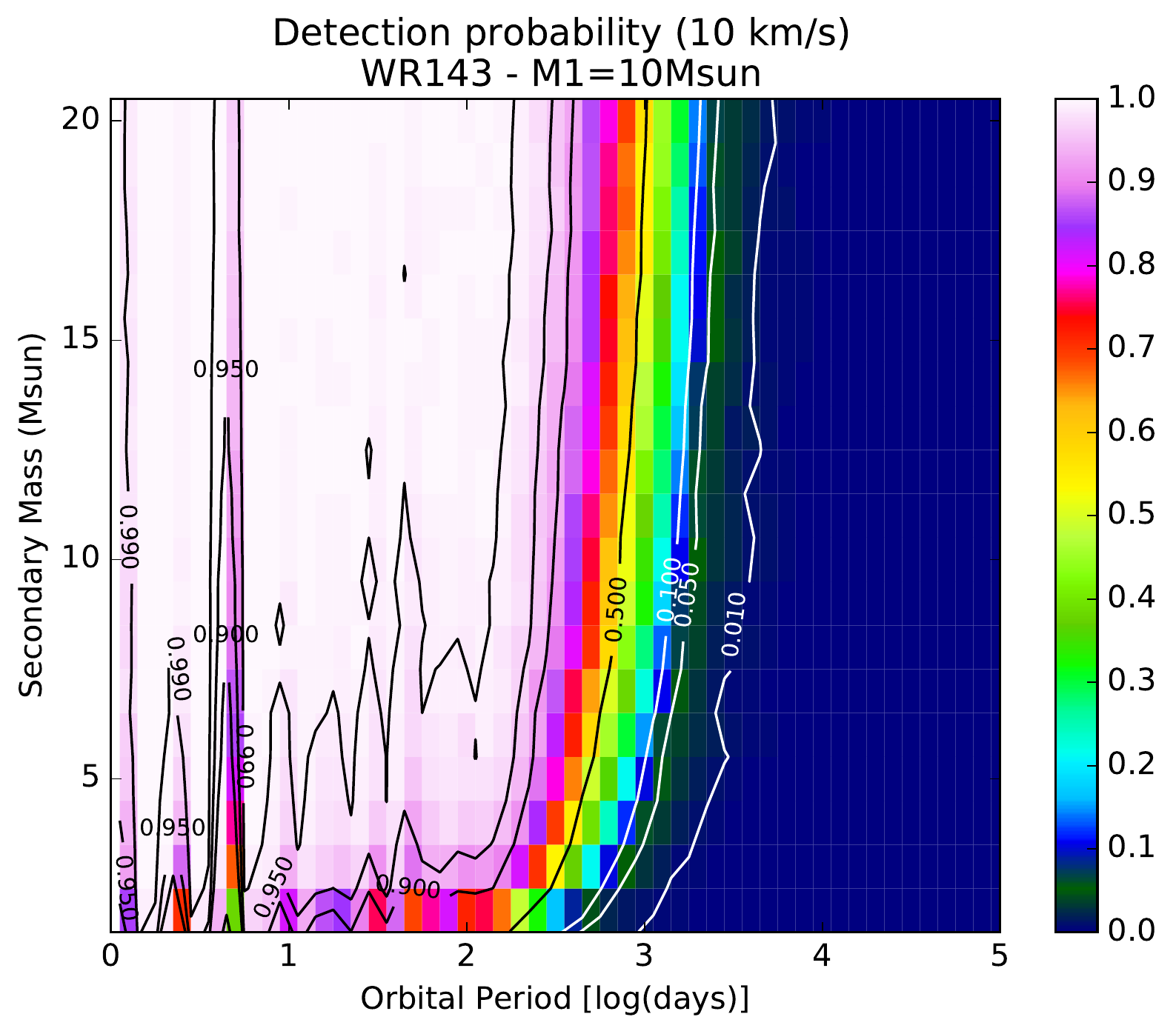}
    \includegraphics[width=8cm]{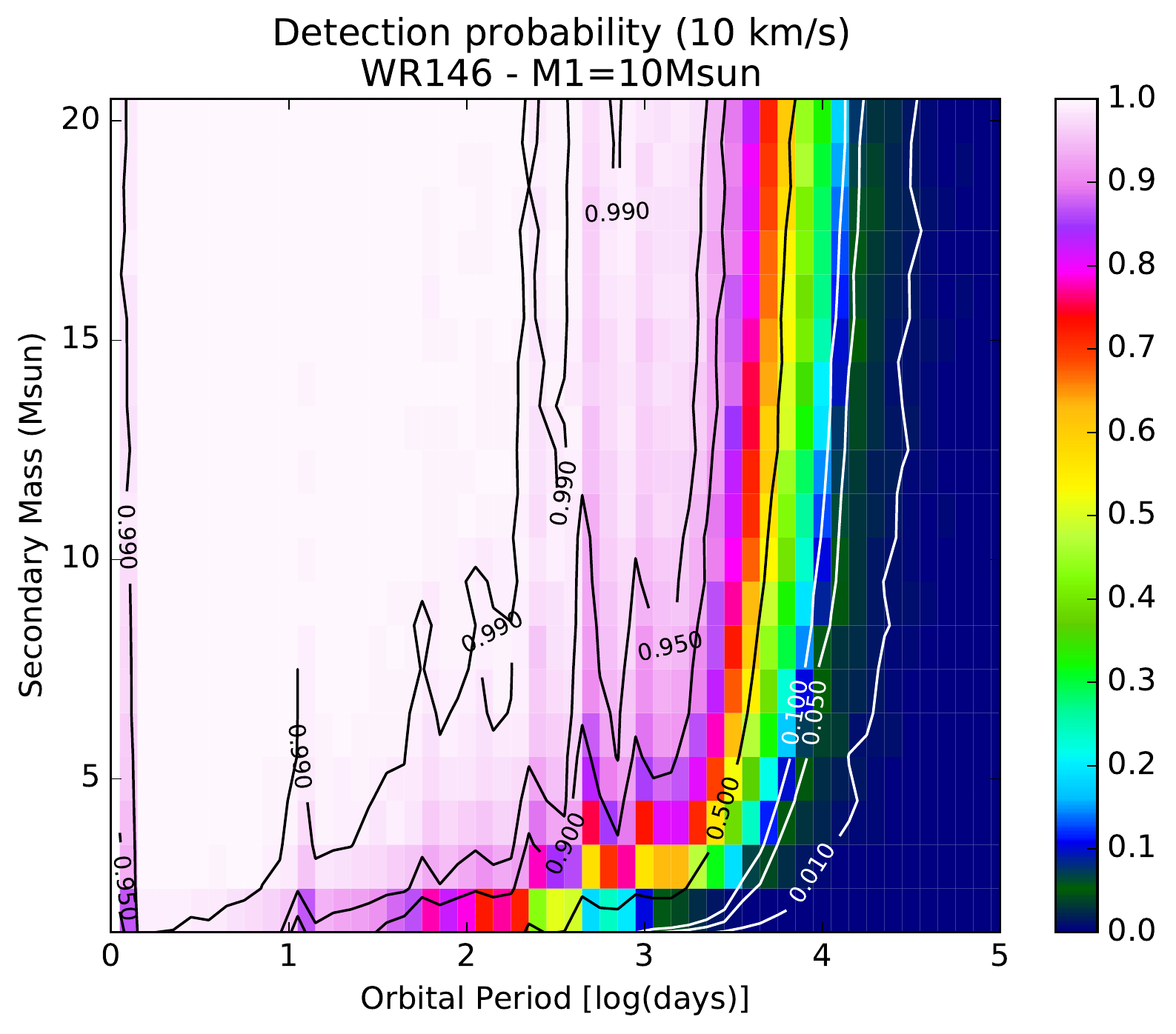}
    \includegraphics[width=8cm]{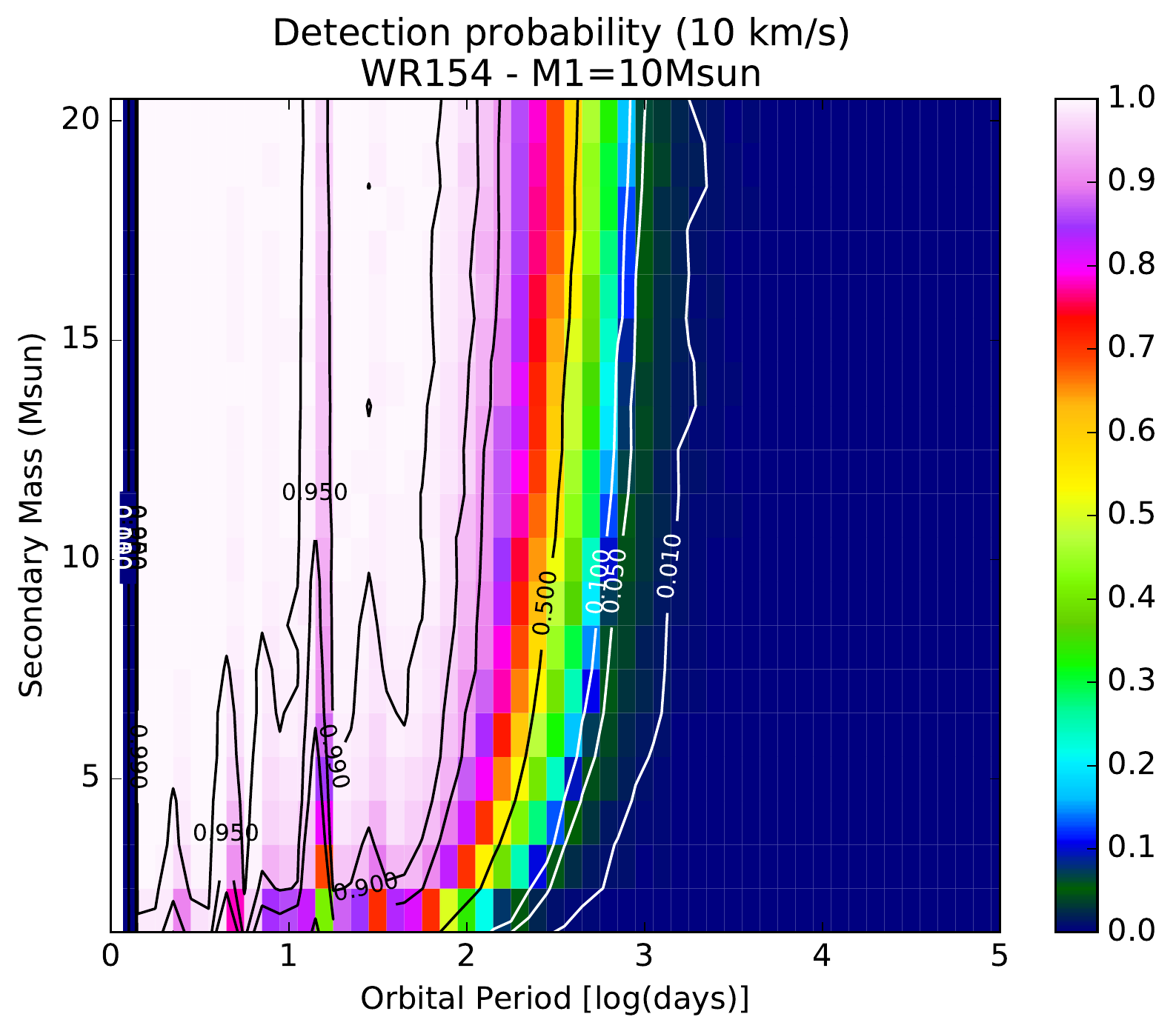}
    \caption{Binary detection probability of the observational campaign for the various stars in our sample, adopting a minimum RV variability threshold of 10~\kms}
    \label{f:Pdetect2}
\end{figure*}
\newpage
\begin{figure*}[h!]
    \centering
    \includegraphics[width=8cm]{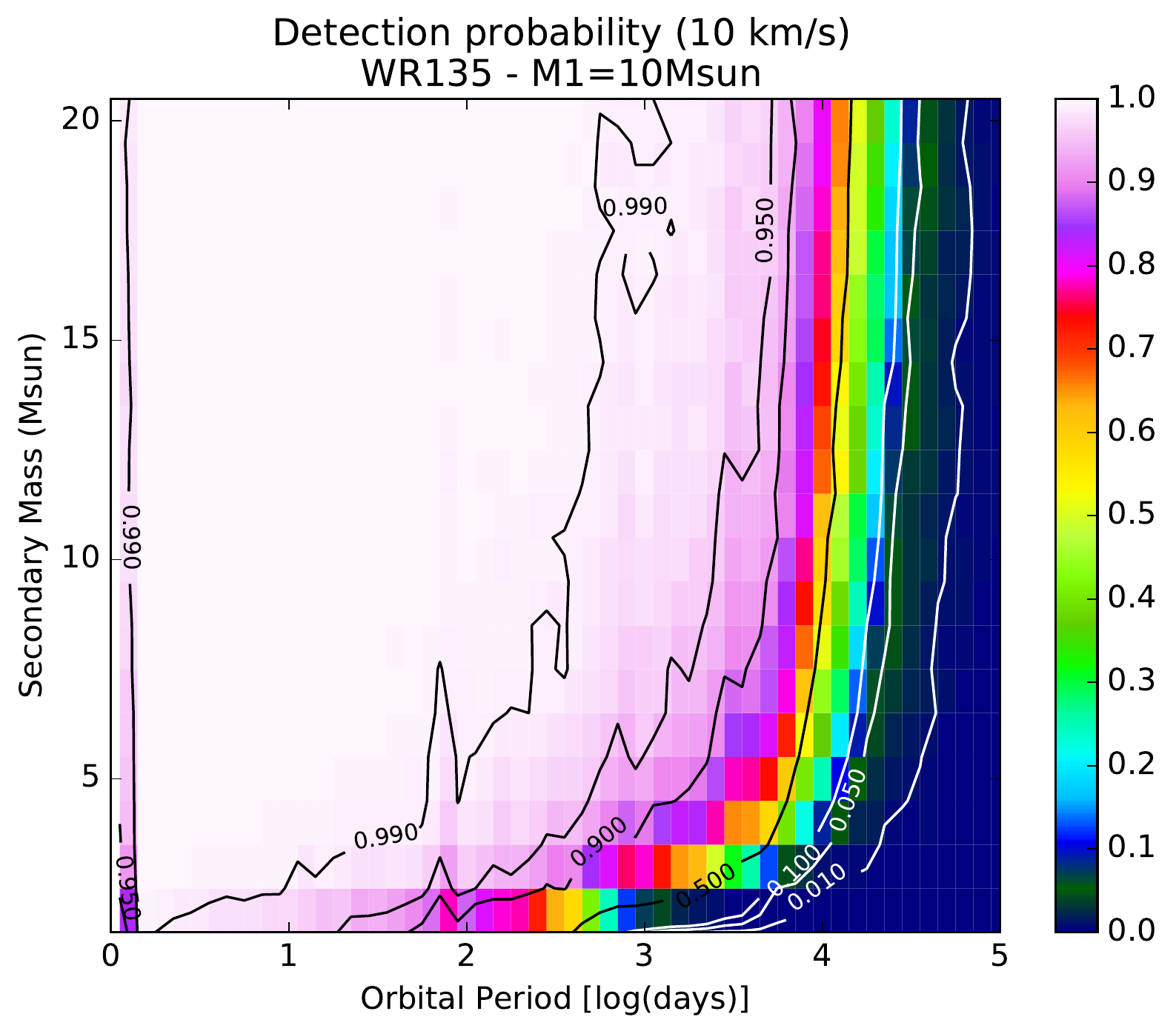}
    \includegraphics[width=8cm]{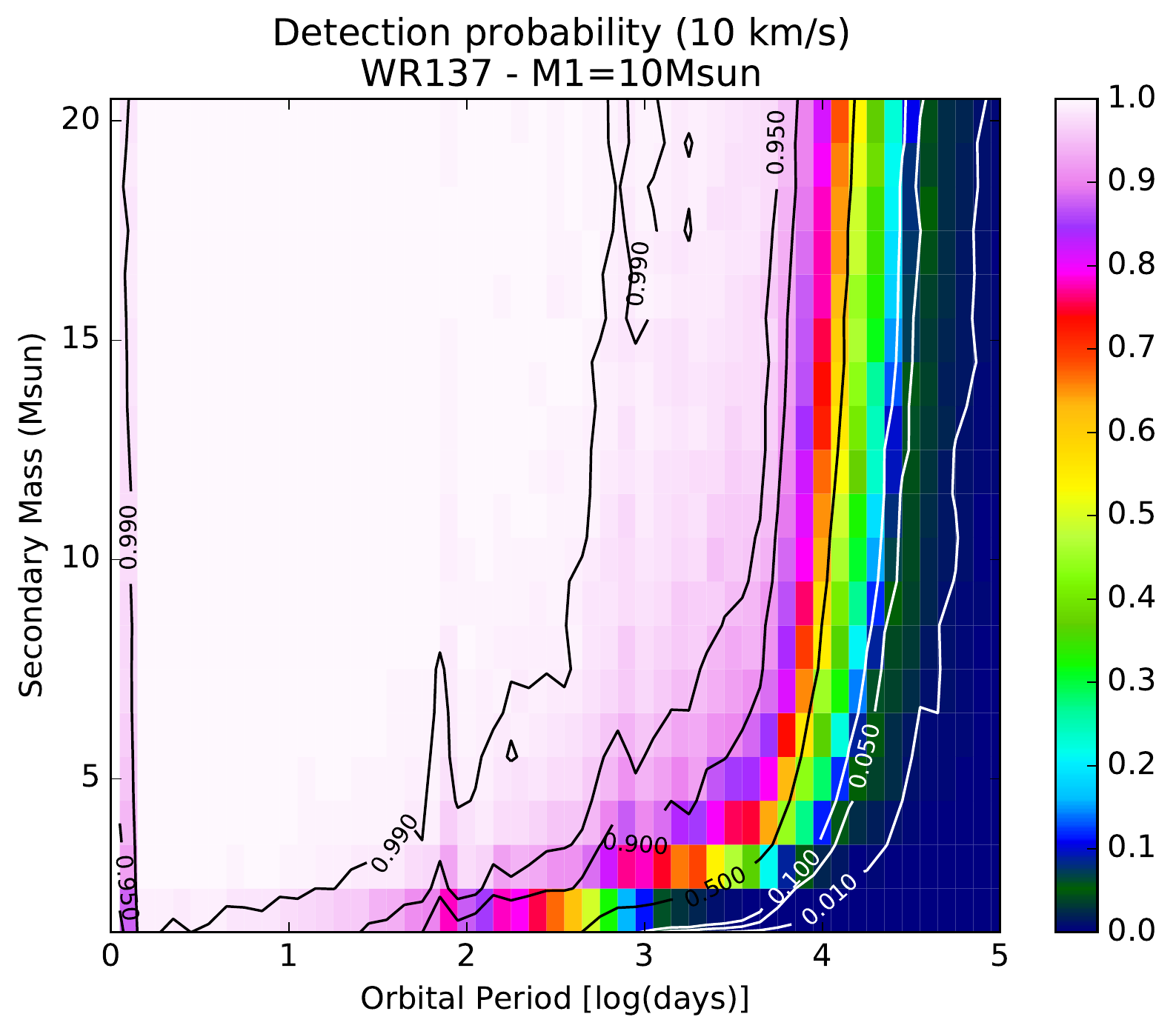}
    \includegraphics[width=8cm]{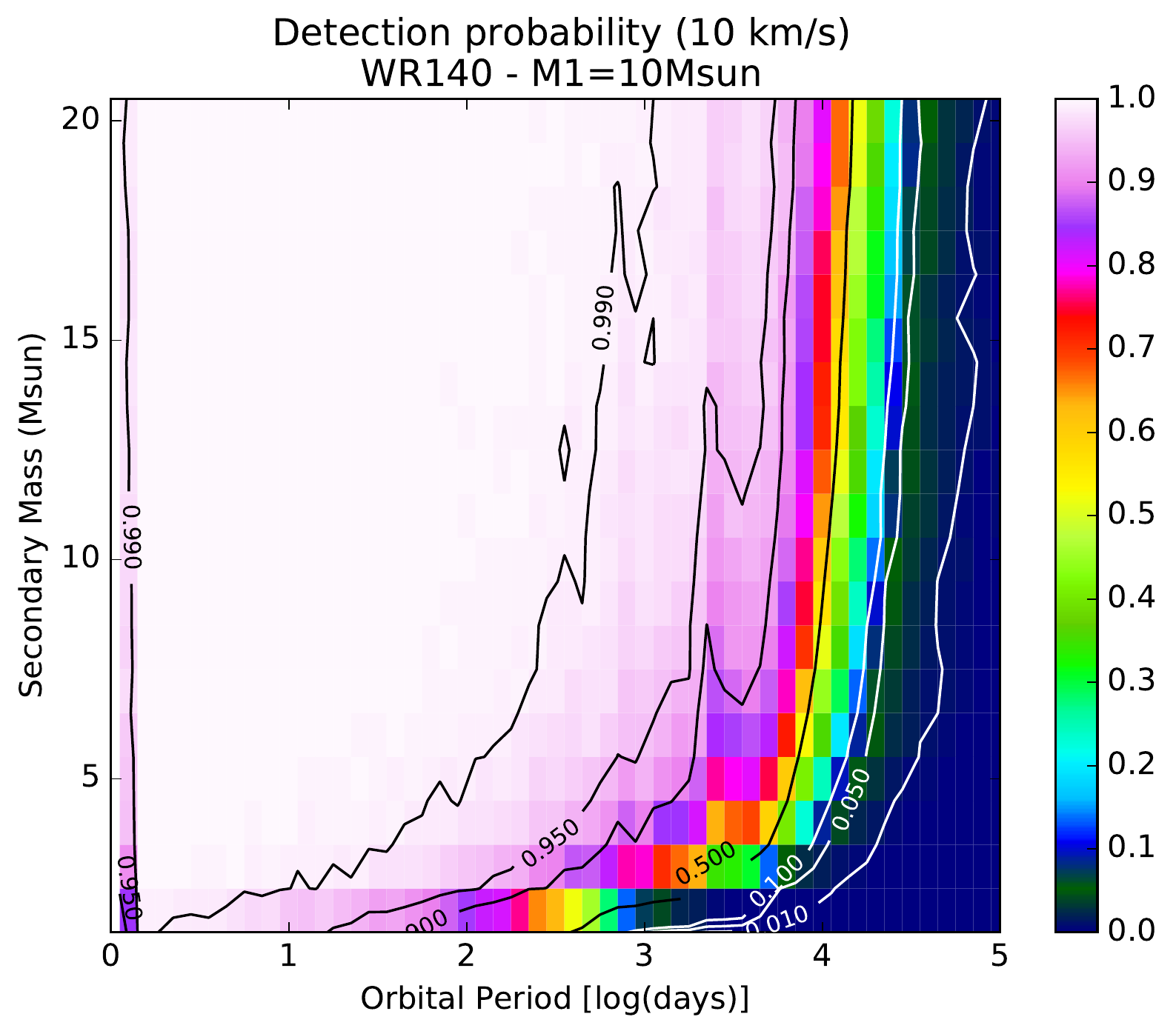}
    \includegraphics[width=8cm]{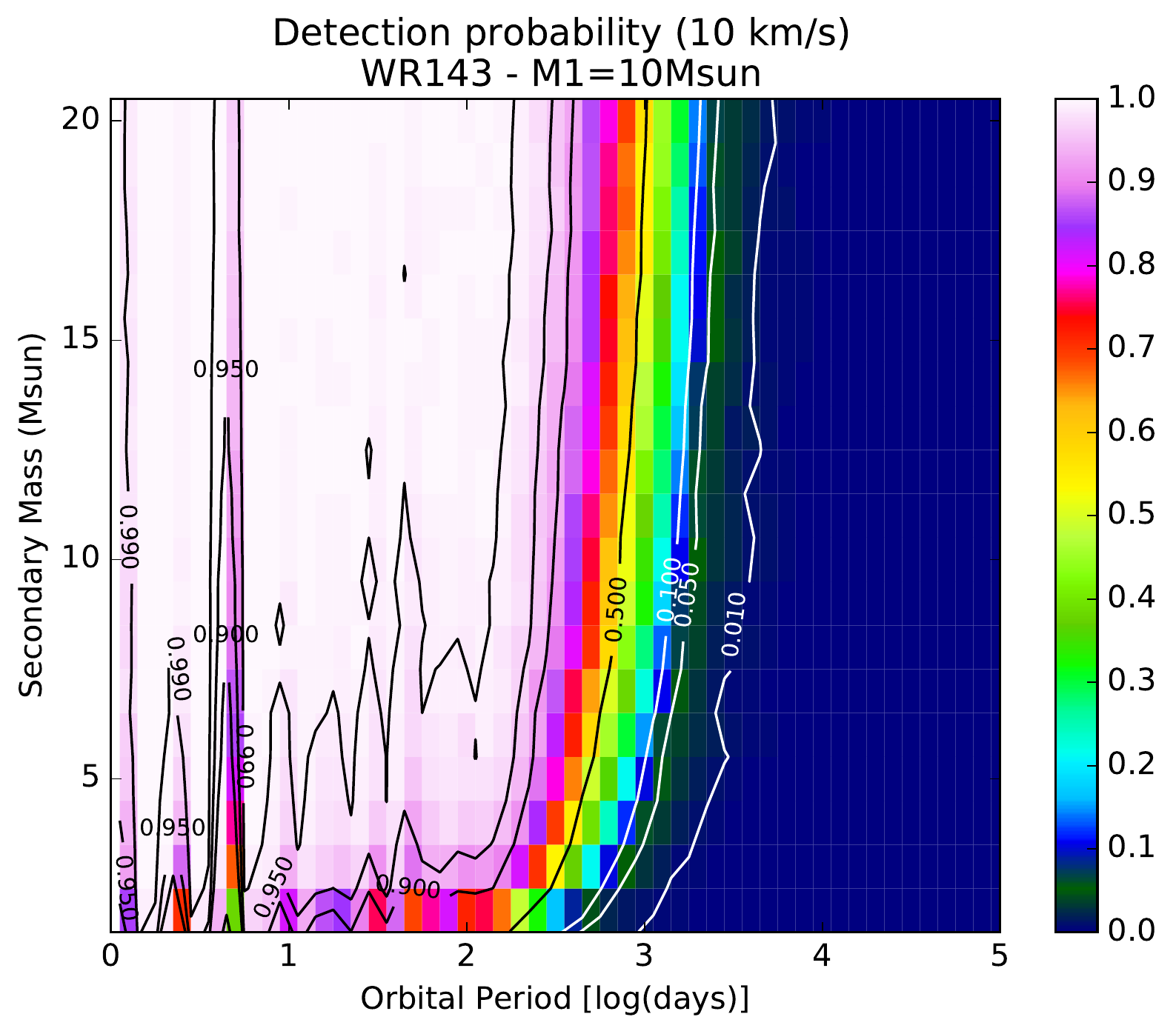}
    \includegraphics[width=8cm]{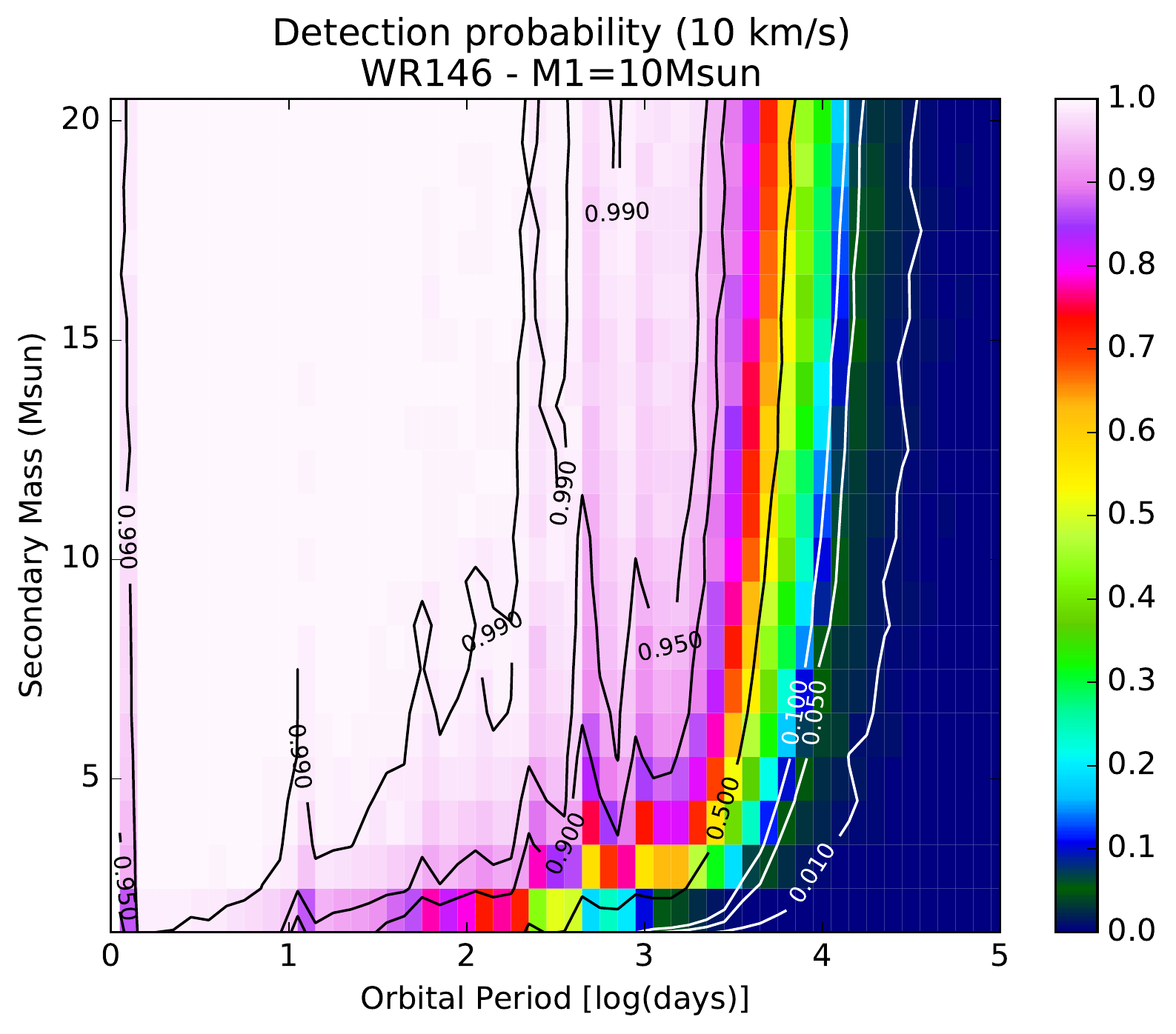}
    \includegraphics[width=8cm]{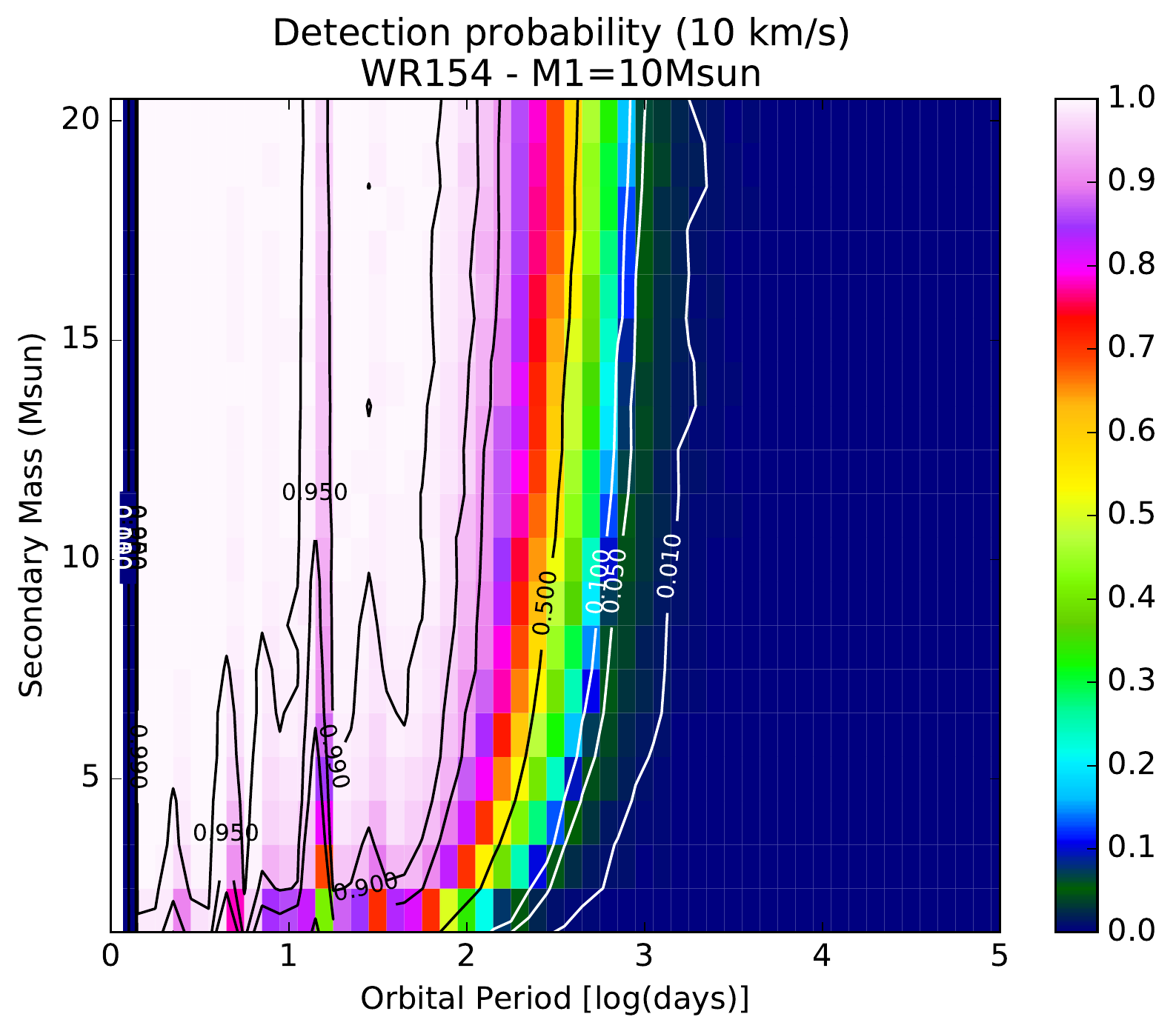}
    \caption{Continued}
    \label{f:Pdetect3}
\end{figure*}
 
% \begin{figure*}
%     \centering
%     \includegraphics[width=\textwidth]{Figures/WR137.pdf}
%     \caption{Phased RV plot for WR 137. Archival RV measurements are shown in blue and are taken from \citet{2005Lefevre}. The RV measurements from HERMES (red) have been shifted by -23.5~\kms in order to convert relative RV measurements to absolute ones. Inset shows the short cadence measurements.}
%     \label{f:WR137_phase}
% \end{figure*}

% \begin{figure*}
%     \centering
%     \includegraphics[width=\textwidth]{Figures/WR140.pdf}
%     \caption{Same as Fig. \ref{f:WR137_phase} but for WR 140. Archival RV measurements are shown in blue \citep{2011Fahed} and green (Thomas et al. submitted). The RV measurements from HERMES (red) have been shifted by 19~\kms in order to convert relative RV measurements to absolute ones.}
%     \label{f:WR140_phase}
% \end{figure*}

% \begin{figure*}
%     \centering
%     \includegraphics[width=\textwidth]{Figures/WR113.pdf}
%     \caption{Same as Fig. \ref{f:WR137_phase} but for WR 113. Archival RV measurements are shown in blue \citep{2018Hill} and green \citep{1981MasseyNiemela}. The RV measurements from HERMES (red and yellow) have been shifted by -135~\kms in order to convert relative RV measurements to absolute ones.}
%     \label{f:WR113_phase}
% \end{figure*}

\end{document}